\documentclass[11pt]{article}
\usepackage{appendix}
\usepackage{amsmath,amssymb,amscd,amsthm,ifthen,color,framed,bm}
\usepackage{graphicx}
\usepackage{rh_defs_21}
\usepackage{listings}

\usepackage{hyperref}
\hypersetup{
    bookmarks=true,         
    unicode=false,          
    pdftoolbar=true,        
    pdfmenubar=true,        
    pdffitwindow=false,     
    pdfstartview={FitH},    
    pdfsubject={Subject},   
    pdfproducer={Producer}, 
    pdfnewwindow=true,      
    colorlinks=true,       
    linkcolor=red,          
    citecolor=blue,        
    filecolor=magenta,      
    urlcolor=cyan           
}
\usepackage{booktabs} 
\usepackage{multirow} 
\usepackage{filecontents}

\usepackage[
backend=bibtex,
url=false,isbn=false,doi=false,
date=year,
hyperref=auto,
style=alphabetic,
natbib=true,
sorting=nty,
firstinits=true,
maxnames=2, maxbibnames=10,
]{biblatex}

\bibliography{references}

\usepackage{pgfplots}
\usepgfplotslibrary{groupplots}
\pgfplotsset{compat=1.18} 

\usetikzlibrary{matrix,arrows,decorations.pathmorphing}

\usepackage{enumitem}

\usetikzlibrary{external}

\usepackage[table]{xcolor}
\usepackage{xparse}
\usepackage{titletoc}
\usepackage{graphicx}
\usepackage[most]{tcolorbox}

\begin{document}

\begin{center}

{
\bf{\LARGE{Piecewise Dynamic Diffusion Regularization for Reconstruction of Cardiac Cine MRI}}
}

\vspace*{.2in}

{\large{
\begin{tabular}{cccc}
Florian Fürnrohr$^{1,2}$, Reinhard Heckel$^{1,2}$ 
\end{tabular}
}}

\vspace*{.05in}

\begin{tabular}{c}
$^1$Technical University of Munich\\ 
$^2$Munich Center for Machine Learning (MCML)
\end{tabular}

\vspace*{.1in}

\today

\vspace*{.1in}

\end{center}


\begin{abstract}
Real-time cardiac cine  MRI enables visualization of the beating heart during free breathing, but severe undersampling and motion make reconstruction highly challenging. 
A central challenge for reconstruction is incorporating powerful priors of cardiac anatomy while remaining computationally efficient. 
We propose \emph{Piecewise Dynamic Diffusion Regularization (PDDR)}, a reconstruction method that integrates a spatiotemporal diffusion model as a generative prior within a variational reconstruction framework for cine MRI. 
The model employs dedicated spatial layers to encode anatomical structure and temporal layers to capture cardiac motion learned from gated cine data. 
PDDR leverages the dynamic prior in a piecewise manner, enabling the efficient use of spatiotemporal diffusion models for processing of long real-time sequences. 
Experiments on retrospectively accelerated and prospective real-time cine MRI demonstrate that PDDR outperforms classical, unsupervised, and diffusion-based methods, delivering high-quality reconstructions with substantially reduced computation time compared to state-of-the-art baselines. These results highlight PDDR as a practical and scalable solution for free-breathing, real-time cardiac MRI. 
Code is available at \url{https://github.com/MLI-lab/pddr}.
\end{abstract}

%
\section{Introduction}
Cardiac cine MRI is an indispensable, non-invasive, clinical imaging technology for evaluation of cardiac function through a video of the beating heart. 

Image reconstruction in cardiac MRI is particularly challenging due to rapid heart motion and limited data acquisition speed in magnetic resonance imaging (MRI). Many conventional techniques are based on binning or gating measurements, and thus implicitly assume each cycle is the same or similar. 
This requires ECG-gated and breathhold acquisitions for temporal binning of the data~\cite{rajiah2023cardiac}.

Real-time cine MRI uses continuously acquired measurements. It enables individual imaging of true physiology under free-breathing conditions. As a result, the acquisition is more efficient, comfortable, and robust~\cite{contijoch2024future}. In real-time cine, acquisition of fully sampled reference data of sufficient spatiotemporal resolution is practically impossible, making video reconstruction exceptionally challenging.  

In real-time acquisitions, only very few measurements are available for each video frame. Reliable reconstruction therefore depends on exploiting the spatiotemporal cardiac structure~\cite{otazo2015low,poddar2015dynamic}. Data-driven approaches can learn problem-specific structure from representative datasets and use this knowledge for improved  reconstruction~\cite{knoll2020deep,heckel2024deep}. Consequently, strong spatiotemporal priors are required for accurate cardiac reconstruction.

Diffusion models provide excellent priors, capturing complex image distributions~\cite{ho2020denoising}. Diffusion-based reconstruction methods have shown strong performance in static MRI~\cite{jalal2021robust}. A spatiotemporal diffusion model could learn the distribution of cardiac motion and provide an effective generative prior for sparsely acquired dynamic measurements. 

To date, high-dimensionality and computationally demanding sampling techniques have hindered widespread application of spatiotemporal diffusion priors for video reconstruction~\cite{daras2024warped,wang2025robust}.
In cardiac MRI, diffusion models have primarily been used for reconstructing short gated acquisitions. 
Long real-time cine sequences are beyond the practical scope of most diffusion-based methods. 

In this work, we propose
\emph{%
\underline{P}iecewise \underline{D}ynamic \underline{D}iffusion \underline{R}egularization
}
for reconstruction of cardiac cine MRI videos. The customized dynamic diffusion prior uses spatial layers to introduce knowledge about cardiovascular anatomy, while temporal layers model the dynamics of cardiac motion and enable exchange of information across the temporal dimension. 
Using a variational approach, the diffusion model can be applied as
piecewise regularizer in real-time cardiac MRI. The framework constitutes a flexible, robust, and computationally efficient reconstruction method that scales from gated cine to long free-breathing acquisitions with hundreds of video frames.

\vspace{2mm}
\noindent 
The main contributions of this paper are:
\begin{itemize}
    \item We propose PDDR, a novel reconstruction method for free-breathing cardiac cine MRI. Through piecewise variational regularization, PDDR enables the efficient use of strong spatiotemporal diffusion priors in long real-time acquisitions.
    \item We show excellent reconstruction quality in retrospective, simulated, and prospective experiments. Across the evaluated settings, PDDR qualitatively and quantitatively matches or outperforms 
    unsupervised baselines as L+S, FMLP, and T-DIP, and diffusion-based methods including DPS and dSTDM.
    \item Piecewise regularization reduces the computational cost of diffusion-based reconstruction while maintaining image quality. Relative to DPS using the same diffusion model, PDDR reduces runtime from 972\,s to 124\,s and GPU memory usage from 41.6\,GB to 11.4\,GB in prospective real-time reconstruction.
\end{itemize}

\section{Background}
Here we provide background on cardiac MRI reconstruction and diffusion models.

\subsection{Cardiac MRI reconstruction problem}
We consider the reconstruction of a complex-valued video consisting of $N$ frames $\mathbf{x} = \left[\mathbf{x}_1, \ldots,  \mathbf{x}_N\right] \in \mathbb{C}^{N \times H \times W} $ from undersampled multi-coil MRI measurements $ \mathbf{y} = \left[\mathbf{y}_1, \ldots,  \mathbf{y}_N\right] \in \mathbb{C}^{N \times C \times L} $ of the beating heart. 
The linear forward model for time frame~$\tau$ and receiver coil $c$ is given as
\begin{equation*}
\mathbf{y}_{\tau, c} = \mathbf{M}_\tau \mathbf{F} \mathbf{S}_c \mathbf{x}_\tau+\mathbf{n}_{\tau, c},
\end{equation*}
where $\mathbf{S}_c$ are coil sensitivity maps, $\mathbf{F}$ is the two-dimensional discrete Fourier transform, $\mathbf{n}_{\tau, c}$ is additive noise, and $\mathbf{M}_\tau$ is a masking operator encoding the sampling pattern.
We stack in the coil dimension~$C$ and define  $\mathbf{A}_\tau = \mathbf{M}_\tau \mathbf{F} \mathbf{S}$, which lets us write the measurement model as 
$\mathbf{y}_\tau = \mathbf{A}_\tau \mathbf{x}_\tau + \mathbf{n}_\tau$.

Cardiac MRI reconstruction is inherently challenging, as slow data acquisition results in only very few measurements corresponding to a single frame $\mathbf{y}_{\tau}$, and cardiac and respiratory motion induce changes across frames.

To alleviate this problem, it is common to perform acquisitions while an electrocardiogram (ECG) is recorded and patients hold their breath, allowing the data to be binned. However, that can lead to binning artifacts and elimination of dynamic variability in the reconstructions~\cite{rajiah2023cardiac}.

In this work, we consider real-time MRI reconstruction without using binning or gating.
Real-time reconstruction enables visualization of true  cardiac physiology in free-breathing acquisition~\cite{contijoch2024future}.

Unsupervised machine learning methods are widely used for real-time cine reconstruction~\cite{yoo2021time,vornehm2025multi,kunz2024implicit,feng2025spatiotemporal}. While these methods flexibly adapt to various acquisition parameters and physiological dynamics, they require long reconstruction times and do not incorporate learned prior information about cardiac images.

Standard supervised techniques are challenging due to the lack of ground truth training data. Existing methods are trained using binned reference acquisitions and a fixed synthetic measurement model, making generalization to real-time acquisitions difficult~\cite{schlemper2017deep,vornehm2025cinevn}.

\subsection{Diffusion models}
\label{sec:diffusion}
Denoising diffusion models~\cite{ho2020denoising,nichol2021improved} define a forward diffusion process, transforming a clean data sample $\mathbf{x}_0 \sim p_{data}$ to standard Gaussian noise $\mathbf{x}_T \sim \mathcal{N}(0, \mathbf{I})$. Each intermediate sample in the diffusion process $t \in [0, T]$ can be described as $\mathbf{x}_t = \sqrt{1-\sigma_t^2}\mathbf{x}_0 + \sigma_t\boldsymbol{\epsilon}$, with $\boldsymbol{\epsilon} \sim \mathcal{N}(0, \mathbf{I})$ and a fixed variance schedule $\sigma_t$.
The reverse process, parameterized by a neural network $\boldsymbol{\epsilon}_\theta (\mathbf{x}_t, t)$ intended to predict the noise $\boldsymbol{\epsilon}$, learns to reverse the diffusion process. Training of the network is done by minimizing the objective $\mathcal{L}(\theta)=\mathbb{E}_{\mathbf{x}_0 \sim p_{data}, t \sim \mathcal{U}(0, T), \boldsymbol{\epsilon} \sim \mathcal{N}(0, \mathbf{I})} \left[ \| \boldsymbol{\epsilon}_\theta (\mathbf{x}_t, t) - \boldsymbol{\epsilon} \|^2_2 \right]$.  Image generation starts from a random Gaussian vector 
and applies iterative denoising to sample 
from the data distribution $p_{data}$.

Diffusion models have been applied successfully in reconstruction of accelerated static MRI~\cite{jalal2021robust,chung2022score,ozturkler2023regularization}, where they serve as robust generative image priors.
Solving inverse problems with diffusion models have predominantly relied on posterior sampling techniques~\cite{song2023pseudoinverse,chung2023dps}. In contrast, variational inference offers a fast and flexible alternative by treating sampling with stochastic optimization. Adopting the diffusion prior as regularizer and enforcing consistency with the measurement data, the regularization by denoising diffusion methodology~\cite{mardani2023variational} formulates image reconstruction as a variational optimization problem. Ozturkler et al.~\cite{ozturkler2023regularization} have applied this to reconstruction of static MRI and report enhanced robustness to distribution shifts and faster sampling rates compared to traditional Langevin-based diffusion sampling. 

\section{Piecewise dynamic diffusion regularization}
We propose a diffusion model-based reconstruction method for cardiac MRI that uses an efficient spatiotemporal diffusion prior as a piecewise regularizer within a variational reconstruction framework. The method, called  
Piecewise Dynamic Diffusion Regularization (PDDR), 
allows flexible adaptation to various acquisition lengths, hardware restrictions, and runtime requirements.
\subsection{Piecewise regularization}
In variational reconstruction of a dynamic video $\mathbf{x} \in \mathbb{C}^{N \times H \times W} $, we optimize for consistency with the measurements and consistency with a prior, in our case in form of a diffusion model~$\boldsymbol{\epsilon}_\theta$. 
In real-time cine MRI the acquisition duration is not restricted, and the total number of video frames $N$ can be large. Therefore, computational restrictions inhibit the application of the diffusion model to the full video sequence.

We propose providing a stochastic estimate of the full regularization signal, by applying the dynamic diffusion model~$\boldsymbol{\epsilon}_\theta$ to a random $Q$-sized block of consecutive frames $ \mathbf{x}_\mathbf{q} =  \left[ \mathbf{x}_q, \ldots, \mathbf{x}_{q+Q-1} \right] \in \mathbb{C}^{Q \times H \times W} $ in each optimization step of the reconstruction. 
The variational objective of our method therefore is 
\begin{equation}
\label{eq:obj}
\min _{\mathbf{x}} \sum_{\tau=1}^N\left\|\mathbf{A}_\tau \mathbf{x}_\tau-\mathbf{y}_\tau\right\|_2^2+ \mathbb{E}_{q, t, \boldsymbol{\epsilon}}\left[w_t\left\|\boldsymbol{\epsilon}_\theta\left(\mathbf{x}_{\mathbf{q, t}}, t\right)-\boldsymbol{\epsilon}\right\|_2^2\right],
\end{equation}
where $\boldsymbol{\epsilon} \sim \mathcal{N}(0, \boldsymbol{I})$ is Gaussian noise, $t \sim \mathcal{U}(\lbrace 0, \ldots, T \rbrace) $ denotes the diffusion step, and
$q \sim \mathcal{U}(\lbrace 1, \ldots, N-Q\rbrace)$ defines the position of the regularization block~$ \mathbf{x}_\mathbf{q} $ in the video. The diffusion model~$\boldsymbol{\epsilon}_\theta$ is applied to the noisy input block $\mathbf{x}_{\mathbf{q, t}} = \sqrt{1-\sigma_t^2}\mathbf{x}_\mathbf{q} + \sigma_t\boldsymbol{\epsilon}$, and the estimated noise residual is weighted with a time-dependent weight~$w_t = \lambda \frac{\sigma_t}{\sqrt{1-\sigma_t^2}}$, where $\lambda$ is a hyperparameter that balances data consistency and regularization strength. The size~$Q$ of the regularization block is a hyperparameter that balances the memory consumption, the runtime, and regularizing effect during inference.

\begin{figure}[tb]
	\centering 
	\includegraphics[width=1.0\textwidth]{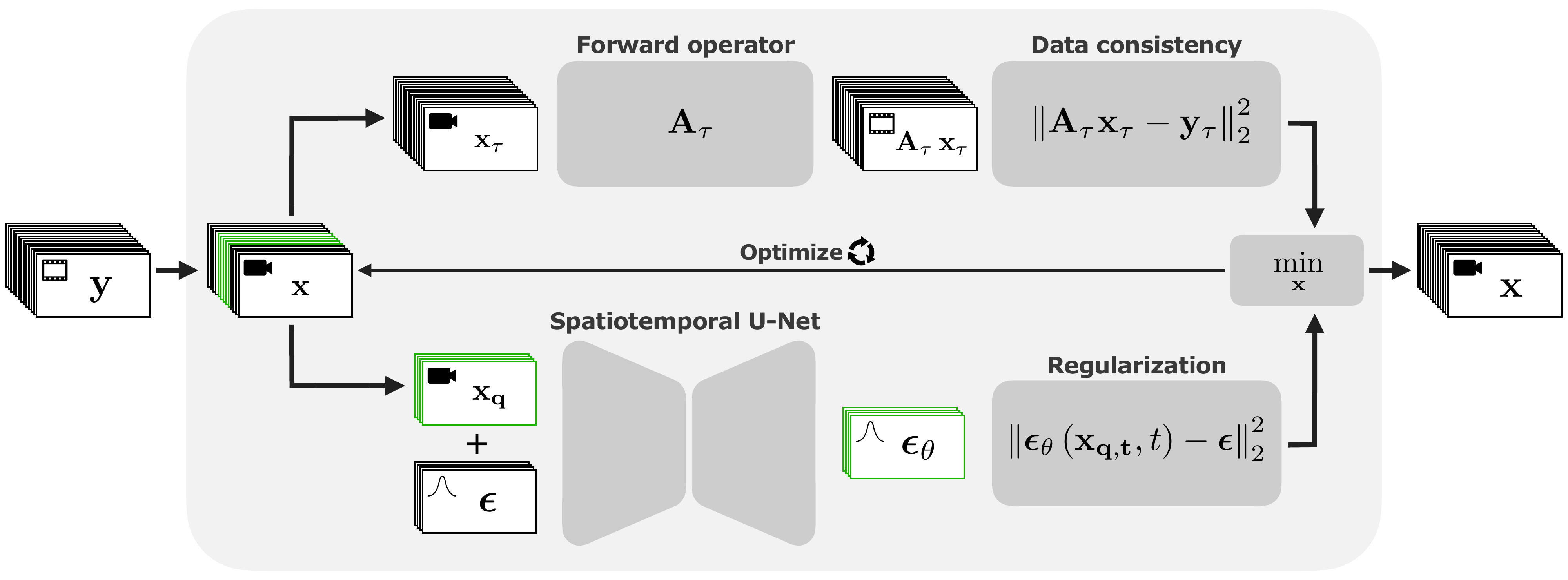}
	\caption{Reconstruction by Piecewise Dynamic Diffusion Regularization. The proposed variational reconstruction uses the full data consistency, but only a subset of frames for regularization in each optimization step.}
	\label{fig:reconstruction}
\end{figure}

The objective~\eqref{eq:obj} is minimized by gradient-based optimization with total number of $K$ steps. In iteration~$k$ of the reconstruction we choose the diffusion step according to $ t = \lfloor T' \cdot \frac{K - k}{K}\rfloor$, with $0 < T' < T$ and $T$ being the number of diffusion steps the model was trained on~\cite{krainovic2024resolution,mardani2023variational}. 
A detailed description of the algorithm choice is given in Appendix~\ref{app: algorithm}.
The reconstruction approach is summarized in Figure~\ref{fig:reconstruction}. 
\subsection{Dynamic diffusion prior}
\label{sec: arch}
As mentioned, for real-time cine MRI reconstruction temporal and spatial relations of the frames have to be leveraged since each frame has only very few measurements associated with it.
Our reconstruction approach is therefore based on an architecture that efficiently learns spatiotemporal relations between frames. 

We propose a separable spatiotemporal U-Net with skippable temporal integration~\cite{blattmann2023align}. As main building block, we introduce a separable spatiotemporal residual block, consisting of a 2D spatial layer and a 1D temporal layer, making it computationally more efficient compared to a naive 3D residual block while still capturing temporal correlations. 
Spatial and temporal layers are convolutional with residual connection, producing outputs $\boldsymbol{\hat{x}}_s$ and $\boldsymbol{\hat{x}}_t$, respectively. The output of the spatiotemporal residual block is then computed as weighted sum $\boldsymbol{\hat{x}} = \sigma(\alpha)\boldsymbol{\hat{x}}_s + (1-\sigma(\alpha))\boldsymbol{\hat{x}}_t$, based on the learnable weight parameter $\alpha \in \mathbb{R}$ and the sigmoid function~$\sigma: \mathbb{R} \rightarrow [0, 1]$. Optionally, the temporal layer can be bypassed, which enables pretraining of the spatial modules only $\boldsymbol{\hat{x}} = \boldsymbol{\hat{x}}_s$. 

The separable spatiotemporal block is then used to build a U-Net architecture with four levels of spatial downsampling, each using four spatiotemporal residual blocks, and one attention block at the bottleneck. 

We compared this architecture to using naive 3D residual blocks, and found that our architecture requires significantly lower memory. This enables using bigger block sizes~$Q$ for a given GPU, which in turn leads to better performance. A detailed model description and ablation is given in Appendix~\ref{app: model}.

\section{Experiments and results}
We study the reconstruction performance  of the proposed dynamic diffusion prior on retrospectively accelerated cardiac MRI. We achieve better or comparable image quality than unsupervised and diffusion-based baselines at lower computational cost. Furthermore, we analyze the efficiency of piecewise regularization on long sequences, and demonstrate high-quality reconstructions of prospective real-time cine MRI. 

\subsection{Retrospective reconstruction}
\label{sec: retro-experiment}
First, we study the reconstruction performance of our method on retrospectively undersampled gated data and compare to baselines. This provides reference-based quantitative evaluation of image quality in cardiac cine, indicating the potential performance for real-time reconstruction.

\paragraph{Data}
We utilize public cine data from the CMRxRecon challenges 2023 and 2024 \cite{wang2024cmrxrecon2023,wang2025cmrxrecon2024} consisting of 2D+time cardiac k-space measurements from 630 healthy subjects, containing multi-slice short-axis (SAX) and single-slice long-axis views (LAX).
Data was acquired on a 3T MAGNETOM Vida scanner (Siemens Healthineers, Germany) in a breath-hold, ECG-gated approach and retrospectively segmented into 12 cardiac phases with a temporal resolution of 50~ms. The spatial resolution was 1.5$\times$1.5~mm$^2$ with a slice thickness of 8.0~mm. The data is provided in a coil compressed format using 10 receiver coil channels. Sensitivity maps were estimated by ESPIRiT~\cite{uecker2014espirit} using the BART toolbox~\cite{uecker2015berkeley}. 

\paragraph{Model training}
\label{sec:trained-model}
We train a dynamic diffusion model~$\boldsymbol{\epsilon}_\theta$ with $T=1000$ steps, 
using 5840 cine sequences from 420 individual subjects of CMRxRecon. 
Training followed the noise-matching methodology described in Section~\ref{sec:diffusion}.

\paragraph{Retrospective undersampling}
For evaluation, we generate variable density $kt$-Gaussian undersampling masks with acceleration factors ranging from 4 to 16 and no explicit fully sampled autocalibration signal (ACS). The masks are applied to 160 unseen test sequences from CMRxRecon and reconstructions are acquired for all tested methods.
For testing reproducibility of the results, we report $mean\pm std$ over 5 runs with different random masks. Furthermore, we ran paired t-tests for statistical analysis, indicating significant differences to PDDR with $p<0.05^*$ and $p<0.01^{**}$.

\paragraph{Baselines}
We compare to the low-rank plus sparse (L+S) reconstruction approach~\cite{otazo2015low}. 
We also compare to two untrained methods for cardiac MRI, Fourier-feature multi-layer perceptrons (FMLP) \cite{kunz2024implicit} and  the time-dependent deep image prior (T-DIP) \cite{yoo2021time}. 
Moreover, we compare to diffusion model-based reconstruction methods.
Diffusion posterior sampling (DPS)~\cite{chung2023dps}, an established inverse problems solver, which we adopt to use the proposed dynamic diffusion prior. 
Dual-directional spatiotemporal diffusion model (dSTDM)~\cite{wang2025robust}, a recent approach for cardiac reconstruction using 2D diffusion priors on the spatiotemporal x-t and y-t image planes. 
Lastly, we propose spatial diffusion regularization (SDR), a variant of our method, using only a frame-based 2D image prior. 

Hyperparameters were determined for each method individually, using a grid search over a validation set and fixed for evaluation on unseen test sets. The exact hyperparameter setting is given in Appendix~\ref{app: hyperparameters}.

\paragraph{Results}
\begin{table}[tb]
    \caption{Reconstruction performance for retrospective 12-fold acceleration of gated data. Comparison in terms of image metrics PSNR, SSIM, NMSE and computational requirements as GPU memory (VRAM) and reconstruction time (Time).}
    \label{tab:retro-results}
    \centering
    \begin{tabular*}{\textwidth}{@{\extracolsep{\fill}}l|ccccc@{\hspace{8pt}}}
    \toprule
    \textbf{Method} \; & \textbf{ PSNR [dB] } & \textbf{ SSIM [\%] } & \textbf{ NMSE } & \textbf{ VRAM [GB] } & \textbf{ Time [s] }  \\
    \toprule
    Zero-filled & 20.17{\fontsize{8}{10} $\pm 0.04^{**}$} & 55.50{\fontsize{8}{10} $\pm 0.17^{**}$} & 1.220{\fontsize{8}{10} $\pm 0.014^{**}$} & 0.60{\fontsize{8}{10} $\pm 0.0^{**}$} & 0.015{\fontsize{8}{10} $\pm 0.00^{**}$} \\
    \midrule
    L+S & 31.26{\fontsize{8}{10} $\pm 0.10^{*}$} & 85.44{\fontsize{8}{10} $\pm 0.18^{**}$} & 0.111{\fontsize{8}{10} $\pm 0.002$} & \textbf{1.16}{\fontsize{8}{10} $\pm 0.0^{**}$} & \textbf{3.98}{\fontsize{8}{10} $\pm 0.00^{**}$} \\
    FMLP & 29.55{\fontsize{8}{10} $\pm 0.33^{**}$} & 67.45{\fontsize{8}{10} $\pm 0.87^{**}$} & 0.481{\fontsize{8}{10} $\pm 0.176^{**}$} & 6.67{\fontsize{8}{10} $\pm 0.0^{**}$} & 868.9{\fontsize{8}{10} $\pm 51.3^{**}$} \\
    T-DIP & 29.73{\fontsize{8}{10} $\pm 0.22^{**}$} & 70.70{\fontsize{8}{10} $\pm 0.84^{**}$} & 0.346{\fontsize{8}{10} $\pm 0.256^{*}$} & 2.53{\fontsize{8}{10} $\pm 0.0^{**}$} & 445.5{\fontsize{8}{10} $\pm 8.12^{**}$} \\
    \midrule
    DPS & 32.78{\fontsize{8}{10} $\pm 0.06$} & 87.78{\fontsize{8}{10} $\pm 0.17$} & \textbf{0.069}{\fontsize{8}{10} $\pm 0.002^{*}$} & 16.27{\fontsize{8}{10} $\pm 0.0^{**}$} & 419.0{\fontsize{8}{10} $\pm 0.11^{**}$} \\
    dSTDM & 31.30{\fontsize{8}{10} $\pm 0.28^{**}$} & 79.59{\fontsize{8}{10} $\pm 1.12^{**}$} & 0.130{\fontsize{8}{10} $\pm 0.008^{**}$} & 12.20{\fontsize{8}{10} $\pm 0.0^{**}$} & 79.21{\fontsize{8}{10} $\pm 0.09^{**}$} \\
    \midrule
    SDR & 28.21{\fontsize{8}{10} $\pm 0.11^{**}$} & 79.13{\fontsize{8}{10} $\pm 0.20^{**}$} & 0.240{\fontsize{8}{10} $\pm 0.006^{**}$} & 4.66{\fontsize{8}{10} $\pm 0.0^{**}$} & 14.77{\fontsize{8}{10} $\pm 0.00^{**}$} \\
    PDDR & \textbf{32.84}{\fontsize{8}{10} $\pm 0.08$} & \textbf{88.06}{\fontsize{8}{10} $\pm 0.14$} & 0.097{\fontsize{8}{10} $\pm 0.002$} & 5.42{\fontsize{8}{10} $\pm 0.0$} & 43.18{\fontsize{8}{10} $\pm 0.01$} \\
    \bottomrule
    \end{tabular*}
\end{table}

\begin{figure}[tb]
	\centering 
    \includegraphics[width=1.0\textwidth]{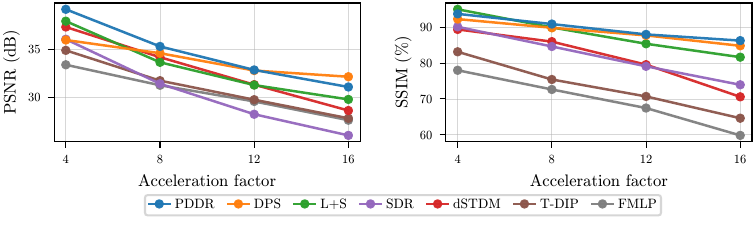}
	\caption{Reconstruction performance for varying levels of undersampling severity. Image quality measured in PSNR~(\textit{left}), and SSIM~(\textit{right}). Performance decreases with increasing acceleration factors.}
	\label{fig:retro-accelerations}
\end{figure}

We find that PDDR achieves high quality reconstructions, outperforming the baselines, as the results reported in Table~\ref{tab:retro-results} show. Here, we focus on results for 12-fold acceleration, matching undersampling patterns required for real-time imaging. 
Results of experiments with accelerations 4, 8, 12, and 16 are given in Figure~\ref{fig:retro-accelerations} and Appendix~\ref{app: experiments}. 

DPS, using the same diffusion model, achieves similar video quality to PDDR but uses a costly sampling approach that requires 3-times more VRAM and about 10-times more reconstruction time compared to PDDR. At lower accelerations DPS underperforms, but at severe undersampling PSNR slightly improves over PDDR. As in real-time cine we are interested in reconstruction of longer sequences, its memory demands make naive DPS sampling highly unpractical.

The diffusion-based cardiac competitor, dSTDM, achieves significantly lower performance than PDDR at significantly higher inference costs. We attribute the comparably low SSIM values of dSTDM to the missing x-y image prior, lacking direct structural image information.

The unsupervised baselines FMLP and T-DIP are designed for scan-specific reconstruction of long prospective measurements and require model fitting during reconstruction, which leads to very high reconstruction times. Both provide comparatively low reconstruction quality in this retrospective setting, with T-DIP being slightly better than FMLP. 
L+S achieves high structural quality for the low acceleration factor 4, but performance declines at the higher accelerations relevant in real-time cine.

Our hypothesis is that for high-performant dynamic MRI, it is important to exploit spatiotemporal correlations in the measurements for reconstruction. Consistent with this hypothesis, the spatial prior does not outperform the classical L+S method which leverages the temporal correlations between frames. This is supported by the reconstruction results using the dynamic diffusion prior in DPS and PDDR, which outperform the spatial prior of SDR in all image metrics by a large margin. This underscores the capabilities of the spatiotemporal diffusion model to capture and utilize strong priors of dynamic cardiac videos.

\subsection{Simulated non-periodic reconstruction}
\label{sec: simulation}
Until now, we only considered gated data, acquired under the assumption of periodic cardiac cycles. In real-time cine imaging, the full dynamic variability over multiple cycles is captured, and periodicity assumptions no longer hold. This is particularly true in free-breathing acquisitions affected by respiratory motion or patients with irregular heartbeats.
Therefore, we study the performance and robustness of the methods in a non-periodic setting through simulation and reconstruction of longer sequences with non-periodic cardiac cycles.
We find that deviations from periodic cardiac dynamics affect all reconstruction methods, particularly those relying on strong temporal signal models. Nevertheless, PDDR consistently achieves high reconstruction quality while exhibiting less performance degradation than methods with stronger low-rank or periodicity assumptions.

\paragraph{Data}
The binned CMRxRecon test data provides short ground truth videos capturing one representative cardiac cycle in $N=12$ video frames. 
Acquired with the assumption of periodically identical cardiac cycles, we simulate periodic acquisitions by repetitive concatenation of the representative cycle. In order to simulate non-periodic videos, we alter the duration of the cardiac cycles by phase-aware interpolation between video frames. To simulate object motion, we optionally add smooth translations. We create simulated videos of $N=60$ frames, approximately covering 5 cardiac cycles, and
synthesize k-space data by application of the forward model. 

%
\paragraph{Simulation settings}
We analyze the reconstruction performance in terms of image quality quantified by PSNR and SSIM. We start by providing reference results for periodic sequences. As real-world cardiac motion is not perfectly periodic, we add mild variations in cycle length to all non-periodic simulations. In order to simulate arrhythmia, we add an irregular heartbeat to the sequence, which consists of a severely faster beat followed by an interval without motion. In free-breathing, respiratory motion further introduces a violation of periodicity, as the object is in different positions during different cardiac cycles. For simplicity, we simulate slow and modest motion of the entire object based on the respiratory rhythm.
Further details are given in Appendix~\ref{app: simulation}.

\paragraph{Results}

\begin{table}[tb]
    \caption{Reconstruction performance for retrospective 12-fold acceleration of simulated data. We simulated perfectly periodic cardiac cycles, severely non-periodic cycles (arrhythmia), and sequences affected by object motion. We analyze image quality in terms of PSNR and SSIM.}
    \label{tab:sim-results}
    \centering
    \begin{tabular*}{\textwidth}{@{\extracolsep{\fill}}l|cc|cc|cc@{\hspace{8pt}}}
    \toprule
    & \multicolumn{2}{c|}{\textbf{Periodic}} & \multicolumn{2}{c|}{\textbf{Arrhythmia}} & \multicolumn{2}{c}{\textbf{Motion}} \\
    \textbf{Method} \; & \textbf{PSNR [dB]} & \textbf{SSIM [\%]} & \textbf{PSNR [dB]} & \textbf{SSIM [\%]} & \textbf{PSNR [dB]} & \textbf{ SSIM [\%] }  \\
    \toprule
    Zero-filled & 20.24$^{**}$ & 56.37$^{**}$ & 20.32$^{**}$ & 56.34$^{**}$  & 20.34$^{**}$ & 56.34$^{**}$ \\
    \midrule
    L+S & \textbf{39.23}$^{**}$ & 93.45$^{**}$ & 34.58$^{**}$ & 89.39$^{**}$ & 31.60$^{**}$ & 84.03$^{**}$ \\
    FMLP & 31.49$^{**}$ & 75.28$^{**}$ & 31.29$^{**}$ & 71.45$^{**}$ & 30.16$^{**}$ & 72.57$^{**}$ \\
    T-DIP & 38.47 & 89.50$^{**}$ & \textbf{37.86} & 88.99$^{**}$ & \textbf{34.67} & 86.78$^{**}$ \\
    \midrule
    DPS & 34.90$^{**}$ & 90.99$^{**}$ & 33.84$^{**}$ & 89.27$^{**}$ & 29.69$^{**}$ & 82.22$^{**}$ \\
    dSTDM & 31.34$^{**}$ & 81.90$^{**}$ & 31.12$^{**}$ & 80.60$^{**}$ & 30.34$^{**}$ & 78.93$^{**}$ \\
    \midrule
    SDR & 30.99$^{**}$ & 86.18$^{**}$ & 30.84$^{**}$ & 85.72$^{**}$ & 30.86$^{**}$ & 85.81$^{**}$ \\
    PDDR & 37.49 & \textbf{94.63} & 36.86 & \textbf{93.91} & 34.05 & \textbf{91.24} \\
    \bottomrule
    \end{tabular*}
\end{table}

Simulation results in Table~\ref{tab:sim-results} show that the image quality of zero-filled and SDR reconstruction are barely affected by differences in the periodicity of cardiac cycles, as they have no assumptions on the temporal behavior. All other reconstruction methods either assume a particular motion model or learn a temporal prior.

The periodic simulation perfectly satisfies the low-rank assumptions made by L+S, efficiently exploiting measurements across cycles.
Consequently, it achieves the highest PSNR in the periodic setup. Performance drops substantially in the non-periodic simulations, yielding significantly lower PSNR and SSIM compared to PDDR. The additional motion violates the low-rank and sparsity assumptions of L+S even more, leading to a severe degradation of image quality.

Although optimized on a scan-specific basis, the unsupervised methods still impose motion assumptions. FMLP enforces smoothness, so it assumes smooth motion rather than periodicity. Results show, it is not able to effectively exploit the redundant measurements of multiple cycles. For that reason, and consistent with the retrospective results, FMLP performs comparatively bad even in the periodic simulation. 
T-DIP, on the other hand, achieves higher PSNR than PDDR across all simulation scenarios, though the difference is not significant, while yielding significantly lower SSIM. Its underlying signal model assumes periodic temporal dynamics and requires an initial estimate of the number of cardiac cycles. This periodicity bias is violated in case of arrhythmia and motion, leading to a slight decrease in image quality. 

The diffusion-based methods, DPS and dSTDM, produce significantly lower reconstruction quality compared to PDDR. Particularly, DPS has a steep performance decrease under simulated motion, suffering from the introduced distribution shift. Using the same diffusion model, PDDRs performance slightly decreases, but still achieves high image quality. In motion simulations with non-periodic cycles, PDDR outperforms DPS by more than 4\,dB in PSNR and 9\,\% in SSIM, indicating much better robustness in reconstruction of unseen physiologic variations.
Additional results on the non-periodic behavior of PDDR can be found in Appendix~\ref{app: simulation}.

We hypothesized that, at higher accelerations, trained methods with strong diffusion priors would outperform unsupervised baselines, consistent with observations in 2D MRI reconstruction~\cite{jalal2021robust, heckel2024deep} and our retrospective results in Section~\ref{sec: retro-experiment}.
In the simulations, however, the untrained methods L+S and T-DIP achieve image quality comparable to the diffusion-based PDDR reconstruction. In cardiac MRI, longer acquisitions offer high spatiotemporal redundancy due to cyclic nature of cardiac motion. The advantage of diffusion models, providing strong priors in low information regimes, becomes less pronounced. Reconstruction methods with signal models that effectively exploit the redundant information, e.g. by low-rank or periodicity assumptions, are able to achieve good performance.

\subsection{Prospective real-time reconstruction}
Finally, we apply PDDR to prospectively accelerated acquisitions of free-breathing data and analyze the reconstruction performance. We use the pretrained model~$\boldsymbol{\epsilon}_\theta$ from Section~\ref{sec:trained-model} with $K=200$ optimization steps and block size~$Q=36$.

\begin{table}[tb]
    \caption{Reconstruction performance for prospective unbinned data. Comparison in terms of signal-to-error ratio (SER), temporal total variation (TTV), GPU memory demand (VRAM), reconstruction time (Time), and model parameter count.}
    \label{tab:prospective-results}
    \centering
    \begin{tabular*}{\textwidth}{@{\extracolsep{\fill}}l|ccccc@{\hspace{8pt}}}
    \toprule
    \textbf{Method} \; & \textbf{ SER [dB] } & \textbf{ TTV } & \textbf{ VRAM [GB] } & \textbf{ Time [s] } & \textbf{ Parameter } \\
    \toprule
    Zero-filled & 0.04$^{**}$ & 91.6$^{**}$ & 4.6$^{**}$ & 0.13$^{**}$ & 0 \\
    \midrule
    L+S & 16.12 & 18.7$^{**}$ & 7.7$^{**}$ & 27.99$^{**}$ & 0 \\
    FMLP & 14.45$^*$ & 7.65$^{**}$ & 7.6$^{**}$ & 3616$^{**}$ & 1.9M  \\
    T-DIP & 16.03 & 33.7 & 7.6$^{**}$ & 1982$^{**}$ & 7.1M \\
    \midrule
    DPS & 14.17$^{**}$ & 39.0$^{**}$ & 41.6$^{**}$ & 972.5$^{**}$ & 15.0M \\
    dSTDM & 15.62 & 56.2$^{**}$ & 28.5$^{**}$ & 230.8$^{**}$ & 65.3M \\
    \midrule
    SDR & 6.19$^{**}$ & 110.4$^{**}$ & 11.2$^{**}$ & 51.35$^{**}$ & 10.9M \\
    PDDR & 16.05 & 31.4 & 11.4 & 124.5 & 15.0M \\
    \bottomrule
    \end{tabular*}
\end{table}

\paragraph{Data}
For our experiment,  we use 20 prospective real-time acquisitions from the public OCMR dataset~\cite{chen2020ocmr}. SAX and LAX data of 1.5T MAGNETOM Avanto and 3.0T MAGNETOM Prisma (Siemens Healthineers, Germany) scanners is provided in spatial resolution of about 2.2$\times$2.2~mm$^2$, slice thickness 8.0~mm, and temporal resolution of 50~ms. The receiver coils had 18 to 34 channels.

\paragraph{Evaluation metrics}
For unbinned performance evaluation we conduct experiments using hold-out frequencies for consistency of the estimated reconstruction with real measurements. Therefore, we randomly subsample 5\% of measured k-space lines for validation, denoted by $\boldsymbol{y}_{\tau, v}$, and only reconstruct the remaining measurements. The estimated validation lines are then extracted from the predicted reconstructions $\boldsymbol{\hat{x}}_\tau$ by applying an adjusted forward operator $\boldsymbol{A}_{\tau, v}$, selecting only the respective validation lines $\boldsymbol{\hat{y}}_{\tau, v} = \boldsymbol{A}_{\tau, v} \boldsymbol{\hat{x}}_\tau$. 
\newline
As quantitative performance metric we compute the signal-to-error ratio (SER) as
\begin{equation*}
    \operatorname{SER} 
    = 10 \log_{10} \frac{\sum_{\tau=1}^N \|\boldsymbol{y}_{\tau, v}\|^2_2}{\sum_{\tau=1}^N \|\boldsymbol{\hat{y}}_{\tau, v} - \boldsymbol{y}_{\tau, v}\|^2_2}.
\end{equation*}

Furthermore, we compute the temporal total variation (TTV) 
quantifying the motion smoothness. High values of TTV can indicate motion artifacts, while lowest values are observed in videos with no motion. 
\paragraph{Results}

\begin{figure}[tb]
	\centering 
	\includegraphics[width=1.0\textwidth]{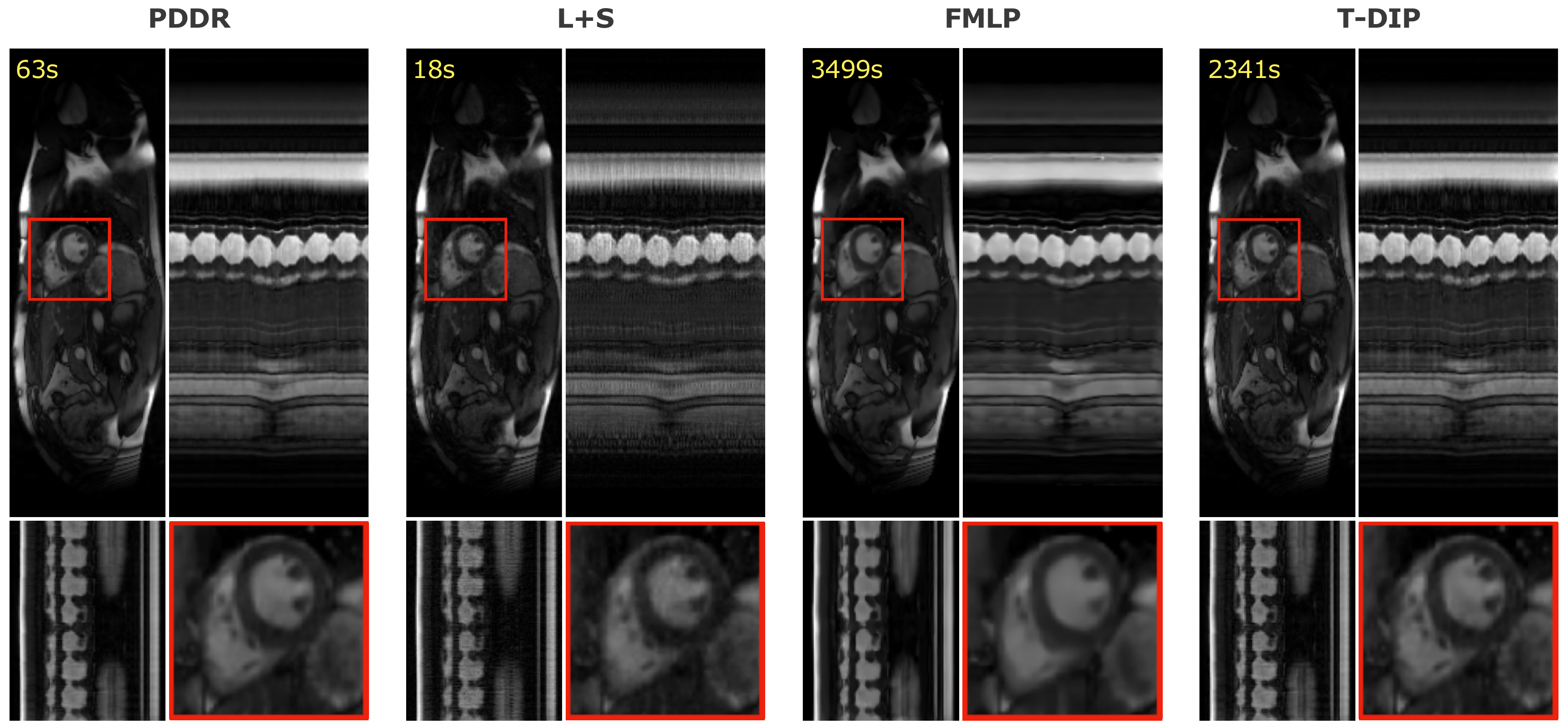}
	\caption{Example reconstructions of prospective free-breathing data. Showing the full field-of-view, time profiles through the heart, and a close-up of the heart. The yellow label indicates the time for reconstruction of the respective video in seconds.}
	\label{fig:real-time_example}
\end{figure}

As shown in Figure~\ref{fig:real-time_example}, PDDR efficiently reconstructs cardiac videos. The figure shows  reconstructions of a free-breathing cine MRI acquisition capturing 7 heartbeats in $N=128$ video frames. 
Quantitative evaluation is presented in Table~\ref{tab:prospective-results}, reporting mean values of reference-free metrics over the prospective testset.

As spatial reference, SDR suffers from strong visual artifacts, testified by low SER and high TTV.
Application of DPS necessitates high GPU memory and runtime, while piecewise regularization with PDDR enables efficient integration of the same dynamic diffusion model for reconstruction while achieving better video reconstruction metrics. 
Reconstruction using the competitive cardiac diffusion-model dSTDM shows significantly higher TTV, VRAM, and reconstruction time, indicating worse quality at higher computational cost.

While classical L+S is the fastest, visual comparison shows strong noise in the reconstructions. Low-rank enforcing can lead to lower motion dynamics of the reconstructions, causing comparably low TTV metrics.
Due to its coordinate-based representation, reconstructions with FMLP are overly smooth and expensive to obtain. T-DIP provides good reconstructions, confirming the simulated results with similar qualitative and quantitative results as PDDR. Nevertheless, T-DIP requires impractical reconstruction times, on average 15-times longer compared to PDDR achieving similar reconstruction quality. 

Although not explicitly represented in the training data of PDDR, we can not observe visual degradation of reconstructions under respiratory motion. In terms of signal estimation capabilities, quantified by SER, we achieve similar or better performance compared to untrained references, showing no performance decline in reconstruction of free-breathing data. 

\subsection{Piecewise regularization ablation}
\label{sec: piecewise}
The proposed piecewise regularization offers a flexible inference procedure by adjustable selection of hyperparameters. Therefore, we analyze PDDR's performance with respect to changes in the reconstruction algorithm, reveling the fundamental tradeoff between computational cost and reconstruction performance. 

\paragraph{Data}
Based on CMRxRecon data we simulate video sequences with up to $N=120$ frames corresponding to 6~s of simulated measurement acquisition. We stick to periodic simulation, to not introduce an additional distribution shift for the ablation. We applied retrospective undersampling masks with acceleration factors 8 and 12.
\paragraph{Results}

\begin{figure}[tb]
	\centering 
    \includegraphics[width=\textwidth]{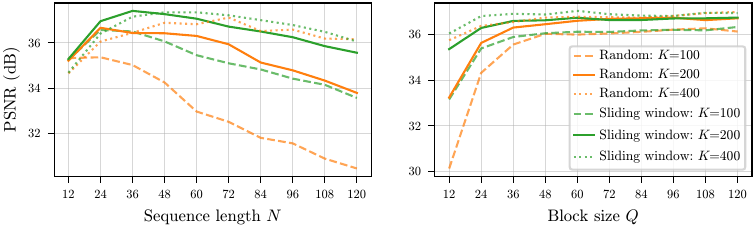}
	\caption{The block sampling strategy. The sliding window approach outperforms random positions of the regularization block. \textit{Left:} For a fixed block size of~$Q=12$, we vary the frame sequence length~$N$. \textit{Right:} For a fixed sequence length~$N=120$, we vary the size of the regularization block~$Q$.}
	\label{fig:block_sampling}
\end{figure}

Figure~\ref{fig:block_sampling} investigates the block sampling strategy and block size~$Q$ with respect to overall sequence length~$N$ and number of optimization steps~$K$. 

Sampling regularization blocks in a sliding window manner outperforms random sampling for the same computation budget, as determined by number of optimization steps~$K$ and block size~$Q$. The smaller the ratio between $Q$ and $N$, the worse random block sampling compares to the sliding window approach. 
The performance difference vanishes with more optimization steps~$K$ or larger block sizes~$Q$, as both leads to more regularization coverage and therefore better approximation of the full regularization signal, but also leads to higher computational cost.

When using sliding window with a moderate number of iterations~$K$, the experiment shows high reconstruction quality even with small block sizes~$Q$ relative to the number of video frames~$N$. This enables the efficient application of otherwise expensive spatiotemporal diffusion models in prospective real-time acquisitions, where regularization of the full sequences is prohibited by memory and runtime constraints. The method can flexibly adopt to  user-specific preferences, by trading memory, runtime, and reconstruction performance through selection of block size~$Q$ and optimization steps~$K$.

\begin{figure}[tb]
	\centering 
    \includegraphics[width=1.0\textwidth]{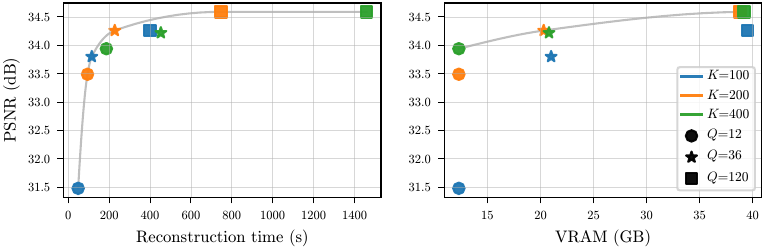}
	\caption{Tradeoff between reconstruction quality and computational cost in PDDR. Colors indicate the number of inference optimization steps~$K$, marker styles the block sizes~$Q$. The estimated pareto-frontier is given in grey. \textit{Left:} Reconstruction time depends on the number of steps~$K$ and block sizes~$Q$. \textit{Right:} The GPU memory demand is independent of $K$, but scales with block sizes~$Q$.}
	\label{fig:computational-tradeoff}
\end{figure}

Figure~\ref{fig:computational-tradeoff} visualizes the tradeoff for a set of reasonable hyperparameters. The block size~$Q$ has a major effect on GPU memory allocation and reconstruction runtime.  We observe a pareto-principle between reconstruction time and image quality, revealing a good performance-computation tradeoff for $K=200$ and $Q=36$. More results on the ablation can be found in the Appendix~\ref{app: experiments}.

\section{Conclusion}
In this paper, we proposed a dynamic diffusion prior as piecewise regularizer in variational reconstruction of cardiac MRI videos.
The method is shown to outperform classical, unsupervised, and diffusion models in reconstruction of accelerated acquisitions. For the more challenging setting of prospective real-time reconstruction, our method yields slightly more accurate reconstructions while being computationally more efficient than current unsupervised and diffusion-based methods. 
Our method is designed to operate on long sequences in a piecewise manner, allowing for efficient handling of extended data while maintaining robust reconstruction quality. This design not only circumvents memory bottlenecks but also offers a principled approach for integrating long-duration, real-time acquisitions into the reconstruction pipeline.
Fast and flexible reconstruction make PDDR a practical approach for cardiac cine, integrating problem-specific spatiotemporal diffusion priors into a highly underdetermined problem.



%
\subsection*{Acknowledgements}
This work is supported by the Munich Center for Machine Learning (MCML), sponsored by the German Federal Ministry of Research, Technology, and Space. Furthermore, we gratefully acknowledge the computational resources provided by the Leibniz Supercomputing Centre (LRZ).

\printbibliography

@article{contijoch2024future,
  title = {The future of CMR: All-in-one vs. real-time CMR (Part 2)},
  author = {Francisco Contijoch and Volker Rasche and Nicole Seiberlich and Dana C. Peters},
  journal = {Journal of Cardiovascular Magnetic Resonance},
  volume = {26},
  number = {1},
  pages = {100998},
  year = {2024},
  issn = {1097-6647},
  xdoi = {10.1016/j.jocmr.2024.100998}
}

@article{rajiah2023cardiac,
  title={Cardiac MRI: state of the art},
  author={Rajiah, Prabhakar Shantha and Fran{\c{c}}ois, Christopher J and Leiner, Tim},
  journal={Radiology},
  volume={307},
  number={3},
  pages={e223008},
  year={2023},
  publisher={Radiological Society of North America}
}

@article{kunz2024implicit,
  author={Kunz, Johannes F. and Ruschke, Stefan and Heckel, Reinhard},
  journal={IEEE Transactions on Computational Imaging}, 
  title={Implicit Neural Networks With Fourier-Feature Inputs for Free-Breathing Cardiac MRI Reconstruction}, 
  year={2024},
  volume={10},
  number={},
  pages={1280-1289},
  xdoi={10.1109/TCI.2024.3452008}
}

@article{vornehm2025multi,
  title = {Multi-dynamic deep image prior for cardiac MRI},
  author = {Vornehm, Marc and Chen, Chong and Sultan, Muhammad A. and Arshad, Syed M. and Han, Yuchi and Knoll, Florian and Ahmad, Rizwan},
  journal = {Magnetic Resonance in Medicine},
  volume = {94},
  number = {6},
  pages = {2668-2679},
  year={2025},
  xdoi = {https://doi.org/10.1002/mrm.70000}
}

@article{yoo2021time,
  title={Time-dependent deep image prior for dynamic MRI},
  author={Yoo, Jaejun and Jin, Kyong Hwan and Gupta, Harshit and Yerly, Jerome and Stuber, Matthias and Unser, Michael},
  journal={IEEE Transactions on Medical Imaging},
  volume={40},
  number={12},
  pages={3337--3348},
  year={2021},
  publisher={IEEE},
  xdoi={10.1109/TMI.2021.3084288}
}

@article{vornehm2025cinevn,
  title={CineVN: Variational network reconstruction for rapid functional cardiac cine MRI},
  author={Vornehm, Marc and Wetzl, Jens and Giese, Daniel and Fürnrohr, Florian and Pang, Jianing and Chow, Kelvin and Gebker, Rolf and Ahmad, Rizwan and Knoll, Florian},
  journal={Magnetic Resonance in Medicine},
  volume={93},
  number={1},
  pages={138--150},
  year={2025},
  publisher={Wiley Online Library},
  xdoi={10.1002/mrm.30260}
}

@article{schlemper2017deep,
  title={A deep cascade of convolutional neural networks for dynamic MR image reconstruction},
  author={Schlemper, Jo and Caballero, Jose and Hajnal, Joseph V and Price, Anthony N and Rueckert, Daniel},
  journal={IEEE Transactions on Medical Imaging},
  volume={37},
  number={2},
  pages={491--503},
  year={2017},
  publisher={IEEE},
  xdoi={10.1109/TMI.2017.2760978}
}

@article{heckel2024deep,
  title={Deep learning for accelerated and robust MRI reconstruction},
  author={Heckel, Reinhard and Jacob, Mathews and Chaudhari, Akshay and Perlman, Or and Shimron, Efrat},
  journal={Magnetic Resonance Materials in Physics, Biology and Medicine},
  volume={37},
  number={3},
  pages={335--368},
  year={2024},
  publisher={Springer}
}

@article{knoll2020deep,
  title={Deep-learning methods for parallel magnetic resonance imaging reconstruction: A survey of the current approaches, trends, and issues},
  author={Knoll, Florian and Hammernik, Kerstin and Zhang, Chi and Moeller, Steen and Pock, Thomas and Sodickson, Daniel K and Akcakaya, Mehmet},
  journal={IEEE Signal Processing Magazine},
  volume={37},
  number={1},
  pages={128--140},
  year={2020},
  publisher={IEEE},
  xdoi={10.1109/MSP.2019.2950640}
}

@inproceedings{jalal2021robust,
  title = {Robust Compressed Sensing MRI with Deep Generative Priors},
  author = {Jalal, Ajil and Arvinte, Marius and Daras, Giannis and Price, Eric and Dimakis, Alexandros G and Tamir, Jon},
  booktitle = {Advances in Neural Information Processing Systems},
  volume = {34},
  year = {2021},
  pages = {14938--14954},
  xeditor = {M. Ranzato and A. Beygelzimer and Y. Dauphin and P.S. Liang and J. Wortman Vaughan},
  xpublisher = {Curran Associates, Inc.},
  xurl = {https://proceedings.neurips.cc/paper_files/paper/2021/file/7d6044e95a16761171b130dcb476a43e-Paper.pdf}
}

@article{chung2022score,
  title={Score-based diffusion models for accelerated MRI},
  author={Chung, Hyungjin and Ye, Jong Chul},
  journal={Medical Image Analysis},
  volume={80},
  pages={102479},
  year={2022},
  publisher={Elsevier},
  xdoi={10.1016/j.media.2022.102479}
}

@inproceedings{blattmann2023align,
  title={Align your latents: High-resolution video synthesis with latent diffusion models},
  author={Blattmann, Andreas and Rombach, Robin and Ling, Huan and Dockhorn, Tim and Kim, Seung Wook and Fidler, Sanja and Kreis, Karsten},
  booktitle={Proceedings of the IEEE/CVF Conference on Computer Vision and Pattern Recognition},
  pages={22563--22575},
  year={2023},
  xdoi={10.1109/CVPR52729.2023.02161}
}

@inproceedings{mardani2023variational,
  title = {A Variational Perspective on Solving Inverse Problems with Diffusion Models},
  author = {Mardani, Morteza and Song, Jiaming and Kautz, Jan and Vahdat, Arash},
  booktitle = {International Conference on Learning Representations},
  year = {2024},
  pages = {28027--28053},
  volume = {2024},
  xeditor = {B. Kim and Y. Yue and S. Chaudhuri and K. Fragkiadaki and M. Khan and Y. Sun},
  xurl = {https://proceedings.iclr.cc/paper_files/paper/2024/file/7711026951f3dc3afb2be427bf98bc73-Paper-Conference.pdf}
}

@inproceedings{ozturkler2023regularization,
  title={Regularization by denoising diffusion process for mri reconstruction},
  author={Ozturkler, Batu and Mardani, Morteza and Vahdat, Arash and Kautz, Jan and Pauly, John M},
  booktitle={NeurIPS 2023 Workshop on Deep Learning and Inverse Problems},
  year={2023}
}

@inproceedings{ho2020denoising,
 author = {Ho, Jonathan and Jain, Ajay and Abbeel, Pieter},
 title = {Denoising Diffusion Probabilistic Models},
 booktitle = {Advances in Neural Information Processing Systems},
 volume = {33},
 year = {2020},
 pages = {6840--6851},
 xpublisher = {Curran Associates, Inc.},
}

@inproceedings{nichol2021improved,
  title={Improved denoising diffusion probabilistic models},
  author={Nichol, Alexander Quinn and Dhariwal, Prafulla},
  booktitle={International Conference on Machine Learning},
  pages={8162--8171},
  year={2021},
  xorganization={PMLR}
}

@inproceedings{chung2023dps, 
 title={Diffusion Posterior Sampling for General Noisy Inverse Problems},          
 author={Hyungjin Chung and Jeongsol Kim and Michael Thompson Mccann and Marc Louis Klasky and Jong Chul Ye},    
 booktitle={International Conference on Learning Representations},           
 year={2023},                
 xurl={https://openreview.net/forum?id=OnD9zGAGT0k}   
}

@inproceedings{song2023pseudoinverse,
  title={Pseudoinverse-guided diffusion models for inverse problems},
  author={Song, Jiaming and Vahdat, Arash and Mardani, Morteza and Kautz, Jan},
  booktitle={International Conference on Learning Representations},
  year={2023}
}

@article{wang2024cmrxrecon2023,
  title={CMRxRecon: A publicly available k-space dataset and benchmark to advance deep learning for cardiac MRI},
  author={Wang, Chengyan and Lyu, Jun and Wang, Shuo and Qin, Chen and Guo, Kunyuan and Zhang, Xinyu and Yu, Xiaotong and Li, Yan and Wang, Fanwen and Jin, Jianhua and others},
  journal={Scientific Data},
  volume={11},
  number={1},
  pages={687},
  year={2024},
  publisher={Nature Publishing Group UK London},
  xdoi={10.1038/s41597-024-03525-4}
}

@article{chen2020ocmr,
  title={OCMR (v1. 0)--open-access multi-coil k-space dataset for cardiovascular magnetic resonance imaging},
  author={Chen, Chong and Liu, Yingmin and Schniter, Philip and Tong, Matthew and Zareba, Karolina and Simonetti, Orlando and Potter, Lee and Ahmad, Rizwan},
  journal={arXiv preprint arXiv:2008.03410},
  year={2020}
}

@article{wang2025cmrxrecon2024,
  title={CMRxRecon2024: A multimodality, multiview k-space dataset boosting universal machine learning for accelerated cardiac mri},
  author={Wang, Zi and Wang, Fanwen and Qin, Chen and Lyu, Jun and Ouyang, Cheng and Wang, Shuo and Li, Yan and Yu, Mengyao and Zhang, Haoyu and Guo, Kunyuan and others},
  journal={Radiology: Artificial Intelligence},
  volume={7},
  number={2},
  pages={e240443},
  year={2025},
  publisher={Radiological Society of North America}
}

@inproceedings{daras2024warped,
  title = {Warped Diffusion: Solving Video Inverse Problems with Image Diffusion Models},
  author = {Daras, Giannis and Nie, Weili and Kreis, Karsten and Dimakis, Alexandros G. and Mardani, Morteza and Kovachki, Nikola B. and Vahdat, Arash},
  booktitle = {Advances in Neural Information Processing Systems},
  volume = {37},
  year = {2024},
  pages = {101116--101143},
  xpublisher = {Curran Associates, Inc.},
  xeditor = {A. Globerson and L. Mackey and D. Belgrave and A. Fan and U. Paquet and J. Tomczak and C. Zhang},
  xurl = {https://proceedings.neurips.cc/paper_files/paper/2024/file/b736c4b0b38876c9249db9bd900c1a86-Paper-Conference.pdf},
  xdoi = {10.52202/079017-3207}
}

@article{poddar2015dynamic,
  title={Dynamic MRI using smoothness regularization on manifolds (SToRM)},
  author={Poddar, Sunrita and Jacob, Mathews},
  journal={IEEE Transactions on Medical Imaging},
  volume={35},
  number={4},
  pages={1106--1115},
  year={2015},
  publisher={IEEE}
}

@article{otazo2015low,
  title={Low-rank plus sparse matrix decomposition for accelerated dynamic MRI with separation of background and dynamic components},
  author={Otazo, Ricardo and Candes, Emmanuel and Sodickson, Daniel K},
  journal={Magnetic Resonance in Medicine},
  volume={73},
  number={3},
  pages={1125--1136},
  year={2015},
  publisher={Wiley Online Library}
}

@article{feng2025spatiotemporal,
  author={Feng, Jie and Feng, Ruimin and Wu, Qing and Shen, Xin and Chen, Lixuan and Li, Xin and Feng, Li and Chen, Jingjia and Zhang, Zhiyong and Liu, Chunlei and Zhang, Yuyao and Wei, Hongjiang},
  journal={IEEE Transactions on Medical Imaging}, 
  title={Spatiotemporal Implicit Neural Representation for Unsupervised Dynamic MRI Reconstruction}, 
  year={2025},
  volume={44},
  number={5},
  pages={2143-2156},
  xdoi={10.1109/TMI.2025.3526452}
}

@article{uecker2014espirit,
  title={ESPIRiT—an eigenvalue approach to autocalibrating parallel MRI: where SENSE meets GRAPPA},
  author={Uecker, Martin and Lai, Peng and Murphy, Mark J and Virtue, Patrick and Elad, Michael and Pauly, John M and Vasanawala, Shreyas S and Lustig, Michael},
  journal={Magnetic Resonance in Medicine},
  volume={71},
  number={3},
  pages={990--1001},
  year={2014},
  publisher={Wiley Online Library}
}

@inproceedings{uecker2015berkeley,
  title={Berkeley advanced reconstruction toolbox},
  author={Uecker, Martin and Ong, Frank and Tamir, Jonathan I and Bahri, Dara and Virtue, Patrick and Cheng, Joseph Y and Zhang, Tao and Lustig, Michael},
  booktitle={Proc. Intl. Soc. Mag. Reson. Med},
  volume={23},
  number={2486},
  pages={9},
  year={2015}
}

@article{krainovic2024resolution,
  title={Resolution-Robust 3D MRI Reconstruction with 2D Diffusion Priors: Diverse-Resolution Training Outperforms Interpolation},
  author={Krainovic, Anselm and Ruschke, Stefan and Heckel, Reinhard},
  journal={arXiv preprint arXiv:2412.18584},
  year={2024}
}

@ARTICLE{wang2025robust,
  title={Robust Cardiac Cine MRI Reconstruction With Spatiotemporal Diffusion Model}, 
  author={Wang, Zi and Huang, Jiahao and Huang, Mingkai and Wang, Chengyan and Yang, Guang and Qu, Xiaobo},
  journal={IEEE Transactions on Computational Imaging}, 
  year={2025},
  volume={11},
  number={},
  pages={1258-1270},
  xdoi={10.1109/TCI.2025.3598421}
}

\newpage

\appendix

\section{Appendix}

\subsection{Model design and ablation}
\label{app: model}
\begin{figure}[]

	\centering 
    \includegraphics[width=0.9\textwidth]{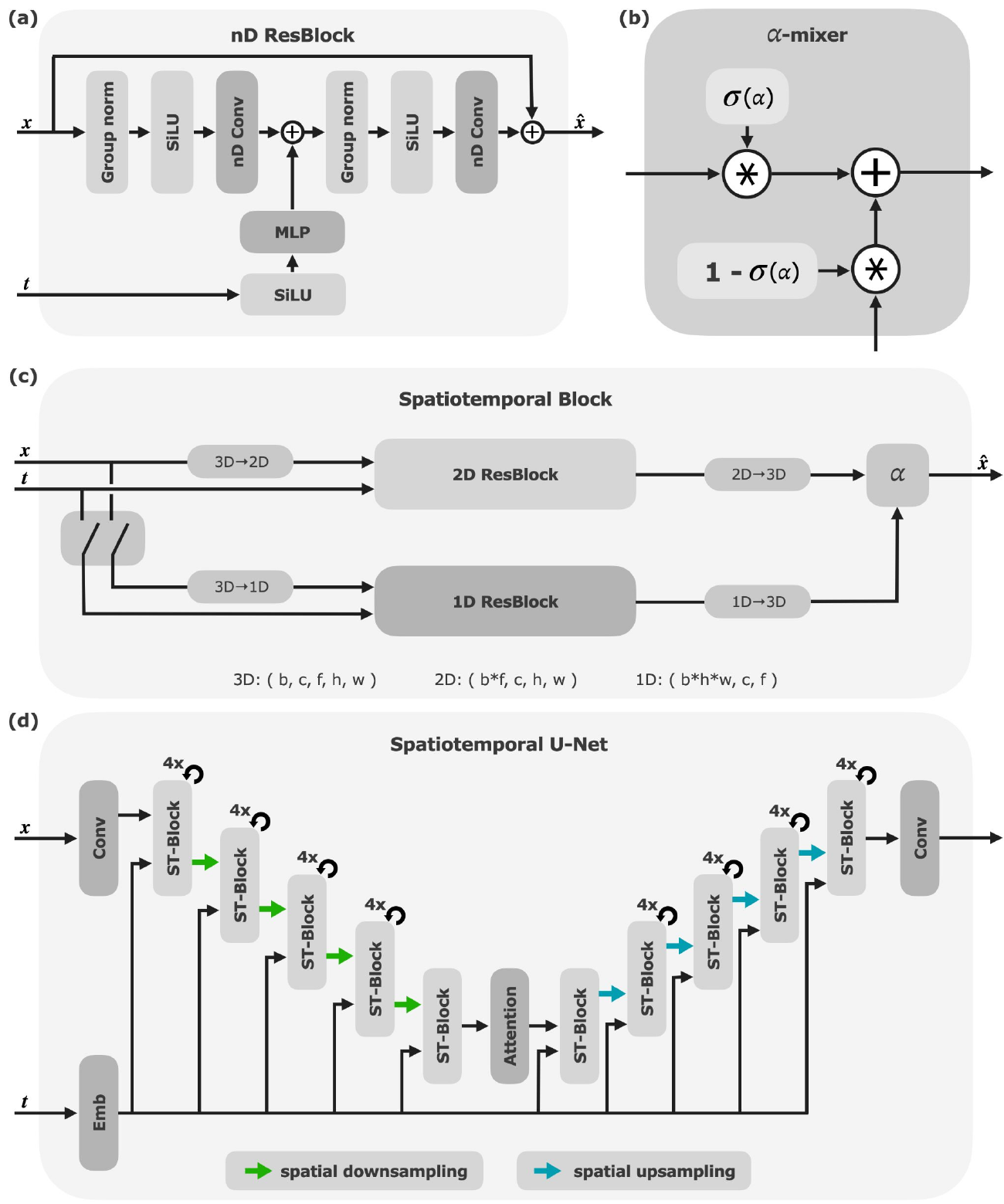}
	\caption{The model architecture. (a) An $n$-dimensional residual block (ResBlock) with timestep conditioning. (b) The $\alpha$-mixer computes a weighted average of the spatial and the temporal layers output with respect to the sigmoid of the learnable parameter ~$\alpha$. (c) The spatiotemporal block (ST-Block) applies a 2D spatial layer and an optional 1D temporal layer to the input and combines the layer outputs. (d) The proposed spatiotemporal U-Net architecture. We use four downsampling levels, four spatiotemporal blocks per level, and an attention layer at the bottleneck.}
	\label{fig:model}

\end{figure}

The proposed model architecture described in Section~\ref{sec: arch} is illustrated in Figure~\ref{fig:model}. The batch of complex-valued input videos, is interpreted as a two-channel real-valued input and given to the model. The model then transforms the input to an initial 32-channel feature space using a 2D convolution. Whenever the U-Net encoder applies downsampling of the spatial dimensions, the channel-dimension is increased using a multiplier with respect to the initial 32-channels, we use precisely $1, 2, 4, 4$. The U-Net decoder, when upsampling spatial dimensions again, transforms the model channels to mirror the encoder channels and concatenates the respective encoder output. An output convolution computes the final output as two-channel video.

\begin{figure}[h]

	\centering 
    \includegraphics[width=1.0\textwidth]{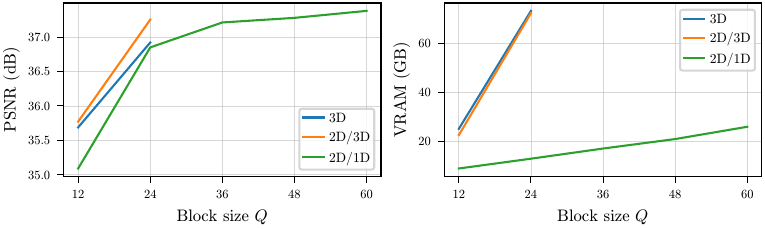}
	\caption{The model ablation. Reconstruction performance measured in PSNR~(\textit{left}), and inference GPU memory consumption ~(\textit{right}), with respect to the regularization block size~$Q$. Longer block sizes increase reconstruction performance, but are prohibited by memory demands when using full 3D architectures.}
	\label{fig:model_ablation}

\end{figure}

As ablation, we compare the proposed architecture with a classical 3D U-Net architecture (3D), which uses naive 3-dimensional residual blocks. As other two architectures, we use the described spatiotemporal block with separated spatial and temporal layers inside the U-Net. The proposed approach uses 1-dimensional temporal layers (2D/1D), while another version uses 3-dimensional temporal layers (2D/3D). We compare reconstruction performance and computational demands over a varying block size~$Q$, using each model architecture. Results are shown in Figure~\ref{fig:model_ablation}. 

For models utilizing 3D convolutions, we were not able to compute results for block sizes larger then 24, as memory demand increases significantly with input size, using already 72~GB of VRAM at block size 24.  Note, that most systems do not have access to GPUs with such large memory during inference. As performance increases with larger batch sizes, and to enable practical use of the method on common GPU hardware, we decided to use 1D temporal layers. 

\clearpage

\subsection{Optimization algorithm details}
\label{app: algorithm}
Reconstruction using the variational approach offers a lot of freedom for choosing inference hyperparameters, as the diffusion model is more a flexible regularizer compared to typical sampling-based approaches. For a diffusion model that was trained with $T=1000$ noising steps, variational inference can perform reconstruction using an independent number of optimization steps~$K$. In each step, one can do regularization using the diffusion model with some random noise level $t \in \lbrace 0, \ldots, T \rbrace$. In their seminal paper, Mardani et al.~\cite{mardani2023variational} already indicate, that a deterministic sampling with decreasing noise levels may increase perceptual reconstruction quality, as similar to standard diffusion sampling reconstructing is performed from coarse structures to fine details. Furthermore, Krainovic et al.~\cite{krainovic2024resolution} show that for reconstruction with random sampling it is not necessary to use all noise levels the generative model was trained for, rather there is an optimal~$T'< T$ for regularizing the reconstruction. We believe the reconstruction does not benefit form high noise levels, as the measurements already provide basic information about the images, in contrast to generation, where we need to generate content from pure noise. Our inference algorithm includes both ideas.

\begin{figure}[h]

	\centering 
    \includegraphics[width=0.85\textwidth]{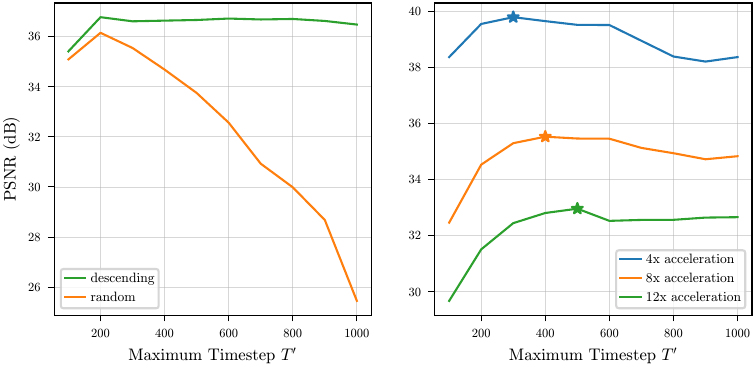}
	\caption{The timestep sampling ablation. Reconstruction performance measured in PSNR with respect to the maximal diffusion timestep~$T'$. \textit{Left:} Uniform random  vs. deterministic descending schedule for sampling of the diffusion steps~$t$ at inference. \textit{Right:} Performance for different levels of undersampling using the descending schedule. Stars indicate the optimal $T'$ for each acceleration factor.}
	\label{fig:timestep_ablation}

\end{figure}

In Figure~\ref{fig:timestep_ablation} one can see the performance difference for decreasing and random timestep sampling using $K=200$. The descending sampling achieves robust, high image quality across the chosen range of noise levels~$t \in \lbrace 0, \ldots, T' \rbrace$ in the diffusion model, whereas the random sampling performs poorly if $T'$ is not optimally chosen. The optimal~$T'$ depends on the acceleration factor of the MRI scan, or in other words on the amount of information provided by the measurements. For higher undersampling factors, less information is provided, and the model benefits more from higher noise levels in the diffusion prior.

\subsection{Implementation details}
\label{app: implementation}
We implemented our method in python using PyTorch. Code to reproduce all results is available at \url{https://github.com/MLI-lab/pddr}. 

Implementations of the baseline methods were taken form public resources, Vornehm et al.~\cite{vornehm2025multi} for L+S\footnote{\url{https://github.com/marcvornehm/M-DIP/blob/main/dip/lps.py}}, Kunz et al.~\cite{kunz2024implicit} for T-DIP\footnote{\url{https://github.com/MLI-lab/cinemri/blob/main/src/models/tdip.py}} and FMLP\footnote{\url{https://github.com/MLI-lab/cinemri/blob/main/src/models/fmlp.py}}, and Chung et al.~\cite{chung2023dps} for DPS\footnote{\url{https://github.com/DPS2022/diffusion-posterior-sampling/blob/main/guided_diffusion/condition_methods.py}}. 
For dSTDM~\cite{wang2025robust}, we implemented the algorithm ourself.
We adjusted all baseline implementations to the used data and measurement parameters. 

For model training, we used 4 NVIDIA H100 GPUs with 94GB of VRAM. Experiments were conducted using a NVIDIA A100 GPU with 40GB/80GB.

\subsection{Hyperparameter setting}
\label{app: hyperparameters}
For reproducibility, we give the exact setting of used reconstruction hyperparameters. Hyperparameters were determined for each method individually,
using a grid search over an validation set. Due to reasons of computational cost, the hyperparameter search was only performed on 12-fold acceleration, not on each acceleration factor individually. This might be a limiting factor for performance, but encourages reproducibility by choosing robust hyperparameters for all settings.

\paragraph{PDDR}
The results presented in Tables \ref{tab:retro-results} and \ref{tab:retro-results-app} were acquired with $K=100$ and 
$Q=N=12$. As in this retrospective setting videos consisted of only 12 frames, taking the piecewise approach is unnecessary and an ablation showed no significant improvement with more optimization steps~$K$, but longer reconstruction time. For long real-time sequences we set $K=200$ and $Q=36$, based on the simulated ablation results reported in Figure~\ref{fig:computational-tradeoff} and Table~\ref{tab:retro-pddr}. We find this to be a good tradeoff between computational cost and reconstruction performance.

Based on the ablation in Figure~\ref{fig:timestep_ablation} and consistent with Krainovic et al.~\cite{krainovic2024resolution}, we choose $T'= 0.4 \cdot T = 400$.
The regularization parameter was set to $\lambda=0.05$, based on a linear grid search on the validation set.

\paragraph{Baselines}
The hyperparameter setting for all baseline methods was determined by ablation over the following parameters. Potential other hyperparameters, e.g. network architecture configurations in untrained methods, were fixed to the values proposed by the original works.

L+S
is a optimization-based method enforcing low-rank and sparsity of the solutions. Therefore, it's performance depends on the maximal number of optimization iterations $K_{max}=600$, the singular-value threshold for low-rank $\lambda_L = 0.1$, and the sparsity threshold $\lambda_S = 0.1$. Early stopping is provided by a relative convergence criterion, i.e. if the difference of consecutive residual norms fall below a threshold.

FMLP and T-DIP
as untrained methods are based on model fitting, the amount of training is a critical aspect. Therefore, the number of training epochs $K$ is a major hyperparameter. Kunz et al.~\cite{kunz2024implicit} provide an early stopping criterion for both methods based on holdout frequencies as validation signal. We choose to set the minimum number of training epochs to $K_{min}=200$, after which we incorporate early stopping by ending the training after the validation loss is not decreasing within $K_{stop}=200$ iterations,  choosing the model with minimum validation error. This approach was more effective than setting a fixed (higher) number of training steps $K$. 

FMLP has additional hyperparameters for spatio-temporal regularization. The spatial $s_x=s_y=120.0$ and temporal $s_t=1.0$ coordinate scales effectively adjust the regularization strength in the respective dimensions. The output scaling $s_{out}$ was highly sensitive to the dataset, so we choose $s_{out}=15.0$ for the retrospective and simulation experiments with CMRxRecon data, and $s_{out}=120.0$ for the prospective evaluations on OCMR.

DPS
is a sampling-based approach, that aligns the sampling process with provided measurements. The hyperparameter $\rho = 40.0$ scales the influence of the data consistency in the sampling, providing the amount of regularization strength. 

SDR,
as a version of our method using a spatial prior, behaves similar to PDDR in the ablations. For fair comparison, we choose $K=100$ and 
$Q=N=12$ as in PDDR. The regularization parameter was also found to perform best at $\lambda=0.05$.

In dSTDM, 
we use the hyperparameter setting suggested by the authors, namely $\rho=1, \lambda=0.5$ and $K=100$, and the exact same model configuration.

\subsection{Additional experimental results}
\label{app: experiments}
\paragraph{Piecewise ablation}
Similar to the ablation provided in Figure~\ref{fig:block_sampling}, we give additional insights on PDDR under different inference settings. Using the same simulation setup as in Section~\ref{sec: piecewise}, Table~\ref{tab:retro-pddr} reports metrics on reconstruction quality and computational cost for different combinations of  $K$ and $Q$, Figure~\ref{fig:computational-tradeoff} visualizes the tradeoff.

As one can see in Figure~\ref{fig:computational-tradeoff}, the block size~$Q$ has a major effect on GPU memory allocation and reconstruction runtime. 
For a given block size, investing more inference time by doing more optimization steps~$K$ can improve image quality, especially when using small block sizes~$Q$. This might be needed for a given consumer GPU. For sufficiently big block sizes~$Q$, using $K=400$ optimization steps does not provide improved reconstruction performance compared to using $K=200$ steps, but approximately doubles reconstruction time.

Using the full regularization signal with $Q=120$ has only minor quality benefits compared to using blocks of size~$Q=36$, but has a major effect on the computational cost of reconstruction. 

Notice, that in this simulation setup, we still were able to perform variational reconstruction on the full signal using a GPU with 40GB VRAM. On the other hand, we were not able to perform DPS sampling for this input size. As mentioned in Section~\ref{sec: retro-experiment}, DPS already uses about 3-times more VRAM as PDDR for the same input size and diffusion model. Piecewise application makes it even more efficient without sacrificing much reconstruction quality. 

\begin{table}[tb]
    \caption{Performance ablation of PDDR using simulation setup with $N=120$ and retrospective undersampling with acceleration factor 12.}
    \label{tab:retro-pddr}
    \centering
    \begin{tabular*}{\textwidth}{c|c|@{\extracolsep{\fill}}ccccc@{\hspace{8pt}}}
    \toprule
    $\mathbf{K}$ & $\mathbf{Q}$ & \textbf{ PSNR [dB] } & \textbf{ SSIM [\%] } & \textbf{ NMSE } & \textbf{ VRAM [GB] } & \textbf{ Time [s] }  \\
    \toprule
     & 12 & 31.48 & 83.29 & 0.100 & 12.3 & 48.31 \\
    100 & 36 & 33.80 & 90.13 & 0.061 & 21.0 & 114.9 \\
     & 120 & 34.26 & 90.53 & 0.054 & 39.5 & 398.9 \\
    \midrule
     & 12 & 33.49 & 89.38 & 0.065 & 12.3 & 94.12 \\
    200 & 36 & 34.26 & 90.56 & 0.057 & 20.3 & 227.2 \\
     & 120 & 34.59 & 90.80 & 0.052 & 38.8 & 746.3 \\
    \midrule
     & 12 & 33.94 & 89.93 & 0.059 & 12.3 & 186.3 \\
    400 & 36 & 34.22 & 90.51 & 0.062 & 20.8 & 452.7 \\
     & 120 & 34.59 & 90.90 & 0.056 & 39.2 & 1459 \\
    \bottomrule
    \end{tabular*}
\end{table}

\paragraph{Experiment with additional ACS}
The results for retrospective undersampling, presented in Tables~\ref{tab:retro-results} and \ref{tab:retro-results-app}, use undersampling masks without any explicit autocalibration signal (ACS). A lot of works on (binned) cardiac cine MRI consider sampling patterns that capture central k-space lines as ACS. Retrospective reconstruction challenges as CMRxRecon additionally add 16 fully sampled lines of the k-space center to the generated undersampling masks~\cite{wang2024cmrxrecon2023,wang2025cmrxrecon2024}. This provides more low-frequency information to the reconstruction and effectively lowers the acceleration factor, making the problem easier to solve. This does not directly relate to our case, as sampling the full center is inapplicable in prospective real-time acquisitions.

To provide a reference for retrospective reconstruction, Table~\ref{tab:retro-results-acs} reports results for 12-fold acceleration with additional 12 lines of ACS. Compared to the results in Table~\ref{tab:retro-results} using no ACS, performance increased for almost all methods, with biggest increases in SDR, dSTDM, and PDDR. Here, PDDR provides significantly better image quality metrics than any other baseline.

\begin{table}[b]
    \caption{Reconstruction performance for retrospective 12-fold acceleration of gated data with 12 central lines of additional ACS.}
    \label{tab:retro-results-acs}
    \centering
    \begin{tabular*}{\textwidth}{@{\extracolsep{\fill}}l|ccccc@{\hspace{8pt}}}
    \toprule
    \textbf{Method} \; & \textbf{ PSNR [dB] } & \textbf{ SSIM [\%] } & \textbf{ NMSE } & \textbf{ VRAM [GB] } & \textbf{ Time [s] }  \\
    \toprule
    Zero-filled & 23.32$^{**}$ & 71.04$^{**}$ & 0.634$^{**}$ & 0.60$^{**}$ & 0.014$^{**}$ \\
    \midrule
    L+S & 34.05$^{**}$ & 91.17$^{**}$ & 0.053 & \textbf{1.16}$^{**}$ & \textbf{3.98}$^{**}$ \\
    FMLP & 30.60$^{**}$ & 70.74$^{**}$ & 0.375$^{**}$ & 6.67$^{**}$ & 819.3$^{**}$ \\
    T-DIP & 31.56$^{**}$ & 82.77$^{**}$ & 0.185$^{*}$ & 2.27$^{**}$ & 829.1$^{**}$ \\
    \midrule
    DPS & 34.62$^{**}$ & 89.55$^{**}$ & 0.044 & 16.27$^{**}$ & 419.0$^{**}$ \\
    dSTDM & 36.00 & 90.03$^{**}$ & 0.044$^{*}$ & 12.2$^{**}$ & 79.4$^{**}$ \\
    \midrule
    SDR & 33.50$^{**}$ & 88.91$^{**}$ & 0.084$^{**}$ & 4.57$^{**}$ & 28.17$^{**}$ \\
    PDDR & \textbf{36.22} & \textbf{92.95} &  \textbf{0.039} & 5.41 & 43.10 \\
    \bottomrule
    \end{tabular*}
\end{table}

\paragraph{Performance over acceleration factors}
The experimental results on retrospective acceleration presented in Table~\ref{tab:retro-results} only included values for 12-fold acceleration. Image quality results for acceleration factors 4, 8, 12, and 16 are given in Table~\ref{tab:retro-results-app}. Computational cost does not depend on the undersampling factor, therefore the same metric results reported in Table~\ref{tab:retro-results} apply.

Figure~\ref{fig:retro-accelerations} provides image quality results with respect to varying undersampling severity. 
For the lower acceleration factor 4, the classical L+S method performs similarly high quality reconstructions, rivaling PDDR which still achieves highest PSNR. In real-time cine MRI, such low acceleration factors are practically not reachable.
For severe undersampling with acceleration factor 16, inference with DPS slightly outperformed the variational reconstruction applied in PDDR in terms of PSNR. Note, that both methods apply the proposed dynamic diffusion prior, DPS only applies a different sampling approach. As mentioned, in real-time cine we typically want to reconstruct longer sequences consisting of much more video frames. DPS utilizes much more inference memory, making naive DPS sampling on larger input sizes impossible.




\begin{table}[h]
    \caption{Reconstruction performance for retrospectively accelerated gated data.}
    \label{tab:retro-results-app}
    \centering
    \begin{tabular*}{\textwidth}{c@{\extracolsep{\fill}}cccccc@{\hspace{8pt}}}
    \toprule
    \textbf{ Acc.} & \textbf{Method} & \textbf{ PSNR [dB] } & \textbf{ SSIM [\%] } & \textbf{ NMSE }  \\
    \toprule
    \multirow{8}{*}{4$\times$}
    & Zero-filled & 25.92{\fontsize{8}{10} $\pm 0.06$} & 70.44{\fontsize{8}{10} $\pm 0.07$} & 0.326{\fontsize{8}{10} $\pm 0.004$} \\
    \addlinespace[1pt]
    \cline{2-5}
    \addlinespace[1pt]
    & L+S & 37.94{\fontsize{8}{10} $\pm 0.05$} & \textbf{95.08}{\fontsize{8}{10} $\pm 0.05$} & \textbf{0.022}{\fontsize{8}{10} $\pm 0.001$} \\
    & FMLP & 33.38{\fontsize{8}{10} $\pm 0.23$} & 78.02{\fontsize{8}{10} $\pm 1.94$} & 0.342{\fontsize{8}{10} $\pm 0.064$} \\
    & T-DIP & 34.89{\fontsize{8}{10} $\pm 0.19$} & 83.19{\fontsize{8}{10} $\pm 1.04$} & 0.050{\fontsize{8}{10} $\pm 0.003$} \\
    \addlinespace[1pt]
    \cline{2-5}
    \addlinespace[1pt]
    & DPS & 35.97{\fontsize{8}{10} $\pm 0.02$} & 92.34{\fontsize{8}{10} $\pm 0.03$} & 0.032{\fontsize{8}{10} $\pm 0.000$} \\
    & dSTDM & 37.33{\fontsize{8}{10} $\pm 0.04$} & 89.45{\fontsize{8}{10} $\pm 0.06$} & 0.044{\fontsize{8}{10} $\pm 0.001$} \\
    \addlinespace[1pt]
    \cline{2-5}
    \addlinespace[1pt]
    & SDR & 36.05{\fontsize{8}{10} $\pm 0.04$} & 90.18{\fontsize{8}{10} $\pm 0.04$} & 0.056{\fontsize{8}{10} $\pm 0.001$} \\
    & PDDR & \textbf{39.18}{\fontsize{8}{10} $\pm 0.05$} & 93.80{\fontsize{8}{10} $\pm 0.05$} & 0.030{\fontsize{8}{10} $\pm 0.001$} \\

    \midrule
    \multirow{8}{*}{8$\times$}
    & Zero-filled & 21.93{\fontsize{8}{10} $\pm 0.04$} & 60.26{\fontsize{8}{10} $\pm 0.10$} & 0.814{\fontsize{8}{10} $\pm 0.010$} \\
    \addlinespace[1pt]
    \cline{2-5}
    \addlinespace[1pt]
    & L+S & 33.64{\fontsize{8}{10} $\pm 0.08$} & 90.01{\fontsize{8}{10} $\pm 0.12$} & 0.066{\fontsize{8}{10} $\pm 0.001$} \\
    & FMLP & 31.24{\fontsize{8}{10} $\pm 0.23$} & 72.65{\fontsize{8}{10} $\pm 0.97$} & 0.390{\fontsize{8}{10} $\pm 0.103$} \\
    & T-DIP & 31.72{\fontsize{8}{10} $\pm 0.11$} & 75.44{\fontsize{8}{10} $\pm 1.48$} & 0.130{\fontsize{8}{10} $\pm 0.009$} \\
    \addlinespace[1pt]
    \cline{2-5}
    \addlinespace[1pt]
    & DPS & 34.59{\fontsize{8}{10} $\pm 0.02$} & 89.91{\fontsize{8}{10} $\pm 0.05$} & \textbf{0.052}{\fontsize{8}{10} $\pm 0.001$} \\
    & dSTDM & 34.17{\fontsize{8}{10} $\pm 0.12$} & 86.03{\fontsize{8}{10} $\pm 0.24$} & 0.080{\fontsize{8}{10} $\pm 0.002$} \\
    \addlinespace[1pt]
    \cline{2-5}
    \addlinespace[1pt]
    & SDR & 31.40{\fontsize{8}{10} $\pm 0.08$} & 84.68{\fontsize{8}{10} $\pm 0.06$} & 0.136{\fontsize{8}{10} $\pm 0.011$} \\
    & PDDR & \textbf{35.30}{\fontsize{8}{10} $\pm 0.07$} & \textbf{90.95}{\fontsize{8}{10} $\pm 0.06$} & 0.062{\fontsize{8}{10} $\pm 0.001$} \\

    \midrule
    \multirow{8}{*}{12$\times$}
    & Zero-filled & 20.17{\fontsize{8}{10} $\pm 0.04$} & 55.50{\fontsize{8}{10} $\pm 0.17$} & 1.220{\fontsize{8}{10} $\pm 0.014$} \\
    \addlinespace[1pt]
    \cline{2-5}
    \addlinespace[1pt]
    & L+S & 31.26{\fontsize{8}{10} $\pm 0.10$} & 85.44{\fontsize{8}{10} $\pm 0.18$} & 0.111{\fontsize{8}{10} $\pm 0.002$} \\
    & FMLP & 29.55{\fontsize{8}{10} $\pm 0.33$} & 67.45{\fontsize{8}{10} $\pm 0.87$} & 0.481{\fontsize{8}{10} $\pm 0.176$} \\
    & T-DIP & 29.73{\fontsize{8}{10} $\pm 0.22$} & 70.70{\fontsize{8}{10} $\pm 0.84$} & 0.346{\fontsize{8}{10} $\pm 0.256$} \\
    \addlinespace[1pt]
    \cline{2-5}
    \addlinespace[1pt]
    & DPS & 32.78{\fontsize{8}{10} $\pm 0.06$} & 87.78{\fontsize{8}{10} $\pm 0.17$} & \textbf{0.069}{\fontsize{8}{10} $\pm 0.002$} \\
    & dSTDM & 31.30{\fontsize{8}{10} $\pm 0.28$} & 79.59{\fontsize{8}{10} $\pm 1.12$} & 0.130{\fontsize{8}{10} $\pm 0.008$} \\
    \addlinespace[1pt]
    \cline{2-5}
    \addlinespace[1pt]
    & SDR & 28.21{\fontsize{8}{10} $\pm 0.11$} & 79.13{\fontsize{8}{10} $\pm 0.20$} & 0.240{\fontsize{8}{10} $\pm 0.006$} \\
    & PDDR & \textbf{32.84}{\fontsize{8}{10} $\pm 0.08$} & \textbf{88.06}{\fontsize{8}{10} $\pm 0.14$} & 0.097{\fontsize{8}{10} $\pm 0.002$} \\

    \midrule
    \multirow{8}{*}{16$\times$}
    & Zero-filled & 19.11{\fontsize{8}{10} $\pm 0.06$} & 52.45{\fontsize{8}{10} $\pm 0.04$} & 1.560{\fontsize{8}{10} $\pm 0.026$} \\
    \addlinespace[1pt]
    \cline{2-5}
    \addlinespace[1pt]
    & L+S & 29.76{\fontsize{8}{10} $\pm 0.09$} & 81.69{\fontsize{8}{10} $\pm 0.25$} & 0.153{\fontsize{8}{10} $\pm 0.004$} \\
    & FMLP & 27.59{\fontsize{8}{10} $\pm 0.41$} & 59.80{\fontsize{8}{10} $\pm 2.15$} & 0.940{\fontsize{8}{10} $\pm 0.375$} \\
    & T-DIP & 27.79{\fontsize{8}{10} $\pm 0.27$} & 64.61{\fontsize{8}{10} $\pm 1.25$} & 0.403{\fontsize{8}{10} $\pm 0.116$} \\
    \addlinespace[1pt]
    \cline{2-5}
    \addlinespace[1pt]
    & DPS & \textbf{32.13}{\fontsize{8}{10} $\pm 0.11$} & 84.87{\fontsize{8}{10} $\pm 0.32$} & \textbf{0.089}{\fontsize{8}{10} $\pm 0.004$} \\
    & dSTDM & 28.60{\fontsize{8}{10} $\pm 0.19$} & 70.59{\fontsize{8}{10} $\pm 1.16$} & 0.218{\fontsize{8}{10} $\pm 0.011$} \\
    \addlinespace[1pt]
    \cline{2-5}
    \addlinespace[1pt]
    & SDR & 26.00{\fontsize{8}{10} $\pm 0.08$} & 73.96{\fontsize{8}{10} $\pm 0.19$} & 0.364{\fontsize{8}{10} $\pm 0.007$} \\
    & PDDR & 31.06{\fontsize{8}{10} $\pm 0.12$} & \textbf{86.32}{\fontsize{8}{10} $\pm 0.10$} & 0.134{\fontsize{8}{10} $\pm 0.004$} \\
    
    \bottomrule
    \end{tabular*}
\end{table}

\begin{figure}[tb]
	\centering 
	\includegraphics[width=1.0\textwidth]{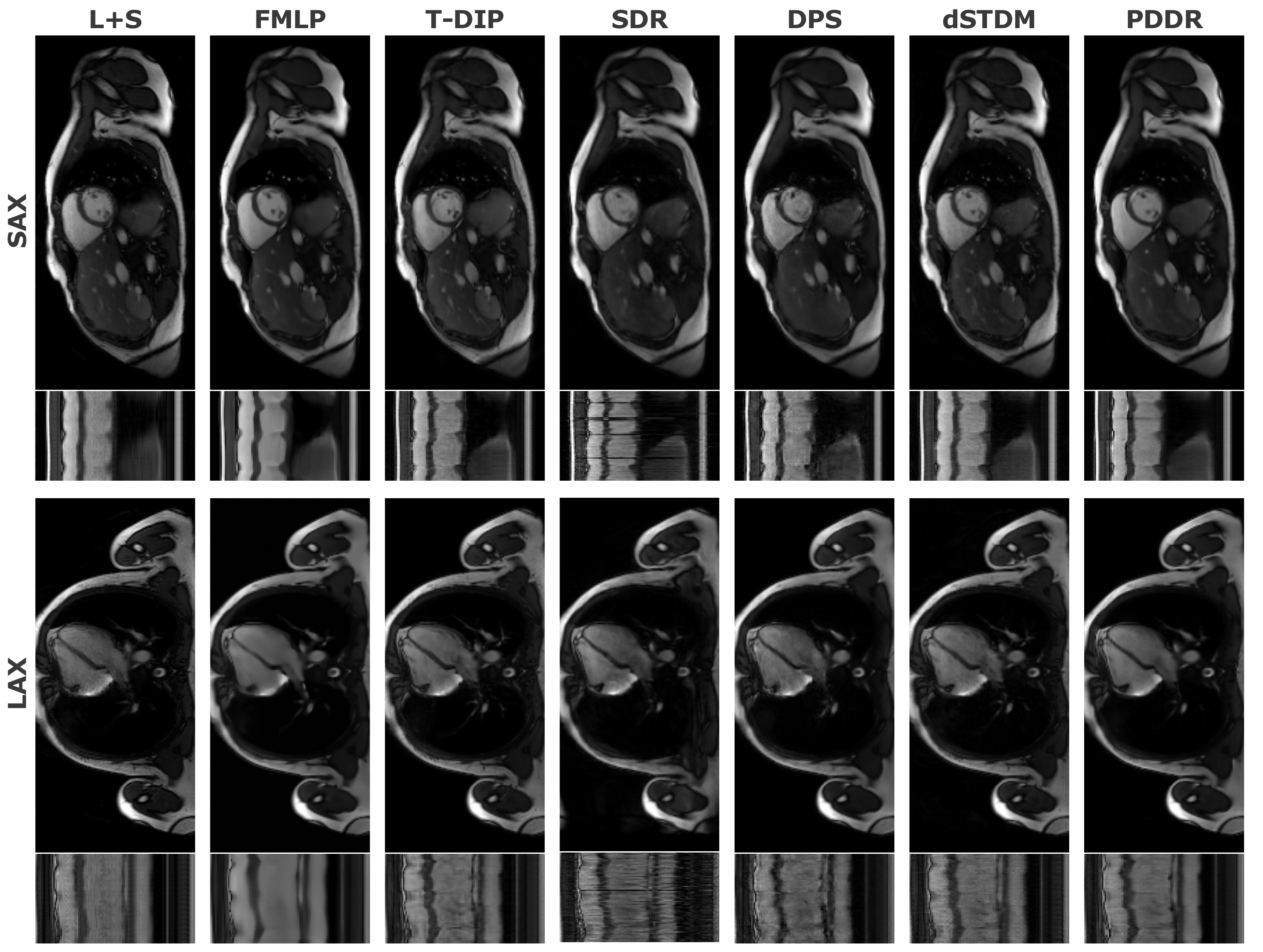}
	\caption{Example reconstructions of prospective SAX and LAX data. Showing the full field-of-view and y-t time profiles through the heart for all considered methods.}
	\label{fig:short_examples}
\end{figure}

\paragraph{Example reconstructions}
Figure~\ref{fig:short_examples} provides additional visual examples of prospective reconstructions. It shows reconstructions of all considered methods for one SAX and one LAX acquisition. 

The visuals align well with the provided qualitative discussion and the quantitative metrics given in Table~\ref{tab:prospective-results}. 
Reconstructions with lowest SER values, SDR and DPS, appear to have degraded image quality, e.g. missing detail in the heart of the SAX reconstructions.
TTV is a good indicator for motion smoothness, observable in the time-profiles of the reconstructions. SDR, as the method with highest TTV, shows strong streaks along time, which appear as intense flickering artifacts in the video. On the other hand, FMLP with significantly lowest TTV looks extremely smooth.

\clearpage
\subsection{Simulation details and results}
\label{app: simulation}
Here we provide additional settings, details, and analysis of non-periodic simulations. Overall, we tested five simulation settings: periodic, mildly non-periodic cycles, arrhythmia, motion, and arrhythmia with motion. 

\paragraph{Simulation details}
For introducing non-periodicity of cardiac cycles, we stretch or compress the duration of the ground truth cycle. For mild non-periodicity, we randomly sample a factor from a Gaussian distribution with mean 1.0 and standard deviation 0.15. For the severely irregular heartbeat, we sample with mean 3.0 and standard deviation 0.5, leading to a cycle that approximately just takes one third of the regular beat, followed by repeating the last frame to create a short post-arrhythmia hold, similar to a skipped beat appearing after a premature heartbeat. The arrhythmia simulation inserts one irregular heartbeat after every 2-3 regular beats, typically leading to one arrhythmic beat in the sequence.

For object motion, we sample a periodic motion trajectory similar to the respiratory cycle. As the respiratory rhythm is way slower than the cardiac cycle, we decided to pick approximately one respiratory cycle into the video consisting of about 5 cardiac beats (Gaussian with mean 1.0 and standard deviation 0.1). We sample motion with maximal x-shift of 1\% of image width, and 0.5\% of image height as maximal y-shift. The motion simulation adds this motion pattern to the mildly non-periodic setting, the arrhythmia with motion simulation adds motion to the arrhythmic case.

\paragraph{Additional results}

\begin{table}[tb]
    \caption{Reconstruction performance for retrospective 12-fold acceleration of simulated data.}
    \label{tab:sim-results-app}
    \centering
    \begin{tabular*}{\textwidth}{@{\extracolsep{\fill}}l|cc@{\hspace{24pt}}|cc@{\hspace{24pt}}}
    \toprule
    & \multicolumn{2}{c|}{\textbf{Non-periodic cycles}} & \multicolumn{2}{c}{\textbf{Arrhythmia+Motion}} \\
    \textbf{Method} \; & \textbf{PSNR [dB]} & \textbf{SSIM [\%]} & \textbf{PSNR [dB]} & \textbf{SSIM [\%]}  \\
    \toprule
    Zero-filled & 20.32$^{**}$ & 56.35$^{**}$ & 20.34$^{**}$ & 56.34$^{**}$ \\
    \midrule
    L+S & 35.21$^{**}$ & 90.09$^{**}$ & 31.47$^{**}$ & 83.83$^{**}$  \\
    FMLP & 31.37$^{**}$ & 72.08$^{**}$ & 30.09$^{**}$ & 70.65$^{**}$ \\
    T-DIP & \textbf{38.15} & 89.01$^{**}$ & \textbf{35.14} & 87.61$^{**}$ \\
    \midrule
    DPS & 33.92$^{**}$ & 89.32$^{**}$ & 29.66$^{**}$ & 82.35$^{**}$ \\
    dSTDM & 31.12$^{**}$ & 80.61$^{**}$ & 30.36$^{**}$ & 78.99$^{**}$ \\
    \midrule
    SDR & 30.87$^{**}$ & 85.77$^{**}$ & 30.84$^{**}$ & 85.82$^{**}$ \\
    PDDR & 36.96 & \textbf{93.96} & 33.95 & \textbf{91.19} \\
    \bottomrule
    \end{tabular*}
\end{table}

\begin{figure}[tb]
	\centering 
	\includegraphics[width=1.0\textwidth]{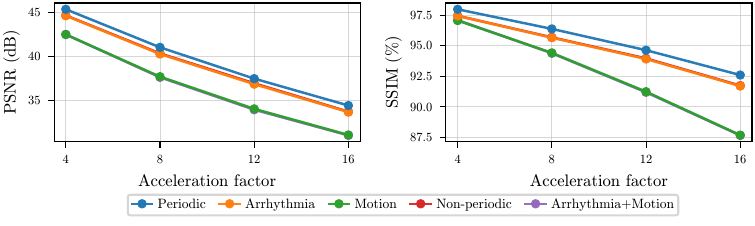}
	\caption{Simulation results of PDDR for multiple accelerations. The performance drop through non-periodic cycles is small and constant across acceleration factors, whereas the drop caused by additional object motion gets more severe at higher accelerations.}
	\label{fig:pddr-simulation-acceleration}
\end{figure}

Table~\ref{tab:sim-results-app} reports the results of the mildly non-periodic and the arrhythmia with motion simulations, not discussed in Section~\ref{sec: simulation}. The results show that the introduction of non-periodic cycles already leads to a small performance drop for all methods. Introducing even more severe non-periodicity by arrhythmia has no observable effect on the image metrics PSNR and SSIM, providing nearly the same results as mild simulations. The simulations with arrhythmia and motion, similarly show that introducing irregular heartbeats has no major effect on the results, as the performance drop mainly is introduced by the motion alone. Figure~\ref{fig:pddr-simulation-acceleration} confirms this observation for PDDR, as differences between arrhythmia and non-periodic, as well as differences between motion and arrhythmia with motion, are unnoticeable. 

Furthermore, we provide additional information about simulated reconstruction performance of PDDR under varying undersampling severity in Figure~\ref{fig:pddr-simulation-acceleration}. We observe really good reconstruction quality for low accelerations with factor 4, declining with increasing problem complexity. Introducing non-periodic cycles, even severe arrhythmic cases, only leads to a small and constant performance drop across accelerations. Introducing motion simulations, results in a higher performance drop, that in terms of SSIM also increases with increasing undersampling. Note, that the simulated motion is more severe than real-world respiratory motion, as it affects the entire object. In free-breathing acquisitions, respiratory motion only affects the heart and inner organs, but the position of the body in the scanner is not affected. Modeling and simulation of realistic motion patterns is challenging, therefore we approximated it with simple translations. We expect, consistent with the prospective experiment,  less severe image degradation in real free-breathing acquisitions.

\end{document}